\documentclass{article}
\usepackage{aas_macros}
\usepackage{natbib}
\usepackage{graphicx}
\usepackage{wasysym}
\def\beq{\begin{equation}}
\def\eeq{\end{equation}}

\def\eref#1{(\ref{eq:#1})}

\def\ie{{i.e.}}
\def\eg{{e.g.}}

\def\bar{\,\rm{bar}}
\def\K{{\,\mbox{K}}}

\def\gcc{\,\rm{g\,cm^{-3}}}

\def\ti#1{$^{{#1}}$}

\def\sec{\,\mbox{s}}
\def\m{\,\mbox{m}}
\def\bfnabla{\mathbf{\nabla}}

\def\req{R_\rm{eq}}

\def\teff{T_\rm{eff}}

\def\eref#1{(\ref{eq:#1})}
\def\rfig#1{fig.~\ref{fig:#1}}

\def\wig#1{\mathrel{\hbox{\hbox to 0pt{%
          \lower.6ex\hbox{$\sim$}\hss}\raise.4ex\hbox{$#1$}}}}

\def\eref#1{(\ref{eq:#1})}
\def\rfig#1{fig.~\ref{fig:#1}}

\def\beq{\begin{equation}}
\def\eeq{\end{equation}}
\def\comment#1{\noindent{\bf }}
\newcommand{\dpar}[2]{{\partial #1\over\partial #2}}

\def\bfnab{{\bf\nabla}}
\def\bfOm{{\bf\Omega}}
\def\bfr{{\bf r}}
\def\ie{{\it i.e.}}

\def\wig#1{\mathrel{\hbox{\hbox to 0pt{%
          \lower.6ex\hbox{$\sim$}\hss}\raise.4ex\hbox{$#1$}}}}

\def\req{R_{\rm eq}}
\def\rpol{R_{\rm pol}}
\def\rbar{{\overline{r}}}

\def\mea{{\rm\,M_\oplus}}

\def\mjup{{\rm\,M_J}}

\def\eg{{\it e.g.}}

\def\teff{T_{\rm eff}}

\def\linc{L_{\rm *p}}

\def\sec{\rm\,s}
\def\m{\rm\, m}
\def\g{\rm\,g}

\def\K{\rm\,K}
\def\bar{\rm\,bar}
\def\gcc{\rm\,g\,cm^{-3}}

\def\p#1{\times 10^{#1}}
\def\ti#1{$^{\rm #1}$}

\def\bfnabla{\mbox{\boldmath $\nabla$}}

\def\apjs{ApJS}
\def\apjl{ApJL}

\def\aap{A\&A}

\def\thebiblio#1{%
 \list{}{\usecounter{dummy}%
         \labelwidth\z@
         \leftmargin 1.5em
         \itemsep \z@
         \itemindent-\leftmargin}
 \reset@font\small
 \parindent\z@
 \parskip\z@ plus .1pt\relax
 \def\newblock{\hskip .11em plus .33em minus .07em}
 \sloppy\clubpenalty4000\widowpenalty4000
 \sfcode`\.=1000\relax
}

\makeatother
\begin{document}
\bibliographystyle{myicarus}

\title{Giant Planets}
\author{Tristan Guillot\\ 
  Observatoire de la C\^ote d'Azur\\
  Laboratoire Cassiop\'ee, CNRS UMR 6202\\
  BP 4229\\
  06304 Nice Cedex 4\\
  France \\
  guillot@obs-nice.fr\\
  \and
  Daniel Gautier\\
  Observatoire de Paris
  LESIA, CNRS FRE 2461\\
  5 pl. J. Janssen\\
  92195 Meudon Cedex\\
  France\\
  gautier@obspm.fr\\
}
\date{To be published in {\em Treatise on Geophysics}, G. Shubert, T. Spohn Eds}

\maketitle

\newpage

\tableofcontents

\newpage

\begin{center}
\Large The Giant Planets

\bigskip
Tristan Guillot\\
Observatoire de la C\^ote d'Azur\bigskip\\
Daniel Gautier\\
Observatoire de Paris\\
\end{center}
\bigskip 

\begin{abstract}
We review the interior structure and evolution of Jupiter, Saturn,
Uranus and Neptune, and extrasolar giant planets with
particular emphasis on constraining their global composition. 
\end{abstract}
%\abstract{
%This is the abstract}

\bigskip

\section{Introduction}

\begin{figure}[htb]
\centerline{\resizebox{12cm}{!}{\includegraphics[angle=0]{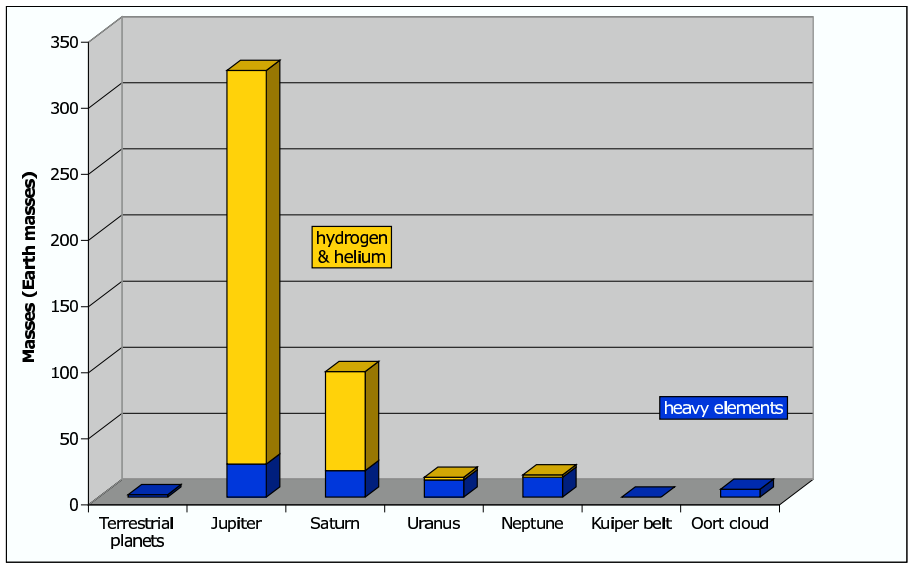}}}
\caption{An inventory of hydrogen and helium and all other elements
  (``heavy elements'') in the Solar System excluding the Sun (the Sun
  has a total mass of $332,960\mea$, including about $5000\mea$ in
  heavy elements, $1\mea$ being the mass of the Earth). The precise
  amount of heavy elements in Jupiter ($10-40\mea$) and Saturn
  ($20-30\mea$) is uncertain (see \S~\ref{sec:JupSat}).}
\label{fig:inventory}
\end{figure}

In our Solar System, four planets stand out for their sheer mass and
size. Jupiter, Saturn, Uranus and Neptune indeed qualify as ``giant
planets'' because they are larger than any terrestrial planet and much
more massive than all other objects in the Solar System, except the
Sun, put together (fig.~\ref{fig:inventory}).  Because of their
gravitational might, they have played a key role in the formation of
the Solar System, tossing around many objects in the system,
preventing the formation of a planet in what is now the asteroid belt,
and directly leading to the formation of the Kuiper belt and Oort
cloud. They also retain some of the gas (in particular hydrogen and
helium) that was present when the Sun and its planets formed and are
thus key witnesses in the search for our origins.

Because of a massive envelope mostly made of hydrogen helium, these
planets are {\it fluid}, with no solid or liquid surface. In terms of
structure and composition, they lie in between stars (gaseous and
mostly made of hydrogen and helium) and smaller terrestrial planets
(solid and liquid and mostly made of heavy elements), with Jupiter and
Saturn being closer to the former and Uranus and Neptune to the
latter. 

The discovery of many extrasolar planets of masses from a few hundreds
down to a few Earth masses and the possibility to characterize them by
the measurement of their mass and size prompts a more general
definition of giant planets. For this review, we will adopt the
following: ``a giant planet is a planet mostly made of hydrogen and
helium and too light to ignite deuterium fusion''. This is purposedly
relatively vague -- depending on whether the inventory is performed by
mass or by atom or molecule, Uranus and Neptune may be included or
left out of the category --. Note that Uranus and Neptune are indeed
relatively different in structure than Jupiter and Saturn and are
generally referred to as ``ice giants'', due to an interior structure
that is consistent with the presence of mostly ``ices'' (a mixture
formed from the condensation in the protoplanetary disk of low
refractivity materials such as H$_2$O, CH$_4$ and NH$_3$, and brought
to the high pressure conditions of planetary interiors --see below).

Globally, this definition encompasses a class of objects that have
similar properties (in particular, a low viscosity and a
non-negligible compressibility) and inherited part of their material
directly from the same reservoir as their parent star. These objects
can thus be largely studied with the same tools, and their formation
is linked to that of their parent star and the fate of the 
circumstellar gaseous disk present around the young star. 

We will hereafter present some of the key data concerning giant
planets in the Solar System and outside. We will then present the
theoretical basis for the study of their structure and evolution. On
this basis, the constraints on their composition will be discussed and
analyzed in terms of consequences for the models of planet
formation.

\section{Observations and global properties}

\subsection{Visual appearances}

\begin{figure}[htb]
%\centerline{\resizebox{12cm}{!}{\includegraphics[angle=0]{inventory.eps}}}
\vspace*{4cm}
\caption{Photographs of Jupiter, Saturn, Uranus and Neptune.}
\label{fig:visual}
\end{figure}

In spite of its smallness, the sample of four giant planets in our
Solar System exhibits a large variety of appearances, shapes, colors,
variability...etc. As shown by fig.~\ref{fig:visual}, all four giant
planets are flattened by rotation and exhibit a more or less clear
zonal wind pattern, but the color of their visible atmosphere is very
different (this is due mostly to minor species in the high planetary
atmosphere), their clouds have different compositions (ammonia for
Jupiter and Saturn, methane for Uranus and Neptune) and depths, and
their global meteorology (number of vortexes, long-lived anticyclones
such as Jupiter's Great Red Spot, presence of planetary-scale storms,
convective activity) is different from one planet to the next.

We can presently only wonder about what is in store for us with
extrasolar giant planets since we cannot image them. But with orbital
distances that can be as close as 0.02\,AU, a variety of masses,
sizes, and parent stars, we should expect to be surprised!

\subsection{Gravity fields}
\label{sec:gravity}

The mass of our giant planets can be obtained with great accuracy from
the observation of the motions of their natural satellites: 317.834,
95.161, 14.538 and 17.148 times the mass of the Earth ($1\mea =
5.97369\times 10^{27}\g$) for Jupiter, Saturn, Uranus and Neptune,
respectively. More precise measurements of their gravity field can be
obtained through the analysis of the trajectories of spacecrafts
during flyby, especially when they come close to the planet and
preferably in a near-polar orbit. The gravitational field thus
measured departs from a purely spherical function due to the planets'
rapid rotation. The measurements are generally expressed by expanding
the components of the gravity field on Legendre polynomials $P_i$ of
progressively higher orders:
\beq
V_{\rm ext}(r,\theta)=-\frac{GM}{r}\left\{
1-\sum_{i=1}^\infty \left(\frac{R_{\rm eq}}{r} \right)^i J_i P_i(\cos\theta)
\right\},
%\label{eq:Vext}
\eeq
where $V_{\rm ext}(r,\theta)$ is the gravity field evaluated outside
the planet at a distance $r$ and colatitude $\theta$, 
$R_{\rm eq}$ is the equatorial radius, and $J_i$ are the
gravitational moments. Because the giant planets are very close to
hydrostatic equilibrium the coefficients of even order are the only
ones that are not negligible. We will see how these gravitational
moments, as listed in table~\ref{tab:moments}, help us constrain the
planets' interior density profiles. 

\begin{table}[htb]
\begin{center}
\caption{Characteristics of the gravity fields and radii}
\label{tab:moments}
\small
\begin{tabular}{l r@{.}l r@{.}l r@{.}l r@{.}l} \hline\hline 
	&\multicolumn{2}{c}{\bf Jupiter} &\multicolumn{2}{c}{\bf
Saturn} &\multicolumn{2}{c}{\bf Uranus} &\multicolumn{2}{c}{\bf
Neptune} \\ \hline
$M\p{-26}$ [kg] & 18&986112(15)\ti{a} & 5&684640(30)\ti{b}
 & 0&8683205(34)\ti{c} & 1&0243542(31)\ti{d} \\
$R_{\rm eq}\p{-7}$ [m] &  7&1492(4)\ti{e} & 6&0268(4)\ti{f} 
 & 2&5559(4)\ti{g} & 2&4766(15)\ti{g} \\
$R_{\rm pol}\p{-7}$ [m] &  6&6854(10)\ti{e} & 5&4364(10)\ti{f} 
 & 2&4973(20)\ti{g} & 2&4342(30)\ti{g} \\
$\overline{R}\p{-7}$ [m] & 6&9894(6)\ti{h} & 5&8210(6)\ti{h} & 2&5364(10)\ti{i} &
  2&4625(20)\ti{i} \\
$\overline{\rho}\p{-3}$ [$\rm kg\,m^{-3}$] & 1&3275(4) & 0&6880(2) & 1&2704(15) & 1&6377(40) \\
$J_2\p{2}$ & 1&4697(1)\ti{a} & 1&632425(27)\ti{b}
 & 0&35160(32)\ti{c} & 0&3539(10)\ti{d}  \\
$J_4\p{4}$ & $-5$&84(5)\ti{a} & $-9$&397(28)\ti{b}
 & $-0$&354(41)\ti{c} & $-0$&28(22)\ti{d} \\
$J_6\p{4}$ & 0&31(20)\ti{a} & 0&867(97)\ti{b}
 & \multicolumn{2}{c}{\dots} & \multicolumn{2}{c}{\dots} \\
$P_{\omega}\p{-4}$ [s] & 3&57297(41)\ti{j} & 3&83624(47)\ ?\ti{j,k}
 & 6&206(4)\ti{l} & 5&800(20)\ti{m} \\
$q$ & 0&08923(5) & 0&15491(10) & 0&02951(5) & 0&02609(23) \\
$C/M\req^2$ & 0&258 & 0&220 & 0&230 & 0&241 \\

\hline\hline
\multicolumn{9}{p{13.5cm}}{%
The numbers in parentheses are the uncertainty in the last digits 
of the given value. The value of the gravitational constant used to
calculate the masses of Jupiter and Saturn is $G=6.67259\p{-11}\,\rm
N\,m^2\,kg^{-2}$ \citep{1987RvMP...59.1121C}. The values of the radii,
density and gravitational moments correspond to the one bar pressure
level (1\,bar$=10^5$\,Pa).}\\
\multicolumn{9}{l}{\ti{a} \citet{1985AJ.....90..364C}}\\
\multicolumn{9}{l}{\ti{b} \citet{2006AJ....132.2520J}}\\
\multicolumn{9}{l}{\ti{c} \citet{1987JGR....9214877A}}\\
\multicolumn{9}{l}{\ti{d} \citet{1989Sci...246.1466T}}\\
\multicolumn{9}{l}{\ti{e} \citet{1981JGR....86.8721L}}\\
\multicolumn{9}{l}{\ti{f} \citet{1985AJ.....90.1136L}}\\
\multicolumn{9}{l}{\ti{g} \citet{1992AJ....103..967L}}\\
\multicolumn{9}{l}{\ti{h} From 4th order figure theory}\\
\multicolumn{9}{l}{\ti{i} $(2\req+\rpol)/3$ (Clairaut's approximation)}\\
\multicolumn{9}{l}{\ti{j} \citet{1986CeMec..39..103D}}\\
\multicolumn{9}{p{13.5cm}}{\ti{k} This measurement from the {\it
    Voyager} era is now 
  in question and values up to $38826$\,s have been proposed (see
  \S~\ref{sec:magnetic fields})}\\ 
\multicolumn{9}{l}{\ti{l} \citet{1986Sci...233..102W}}\\
\multicolumn{9}{l}{\ti{m} \citet{1989Sci...246.1498W}}\\
\end{tabular}
\normalsize
\end{center}
\end{table}

Table~\ref{tab:moments} also indicates the radii obtained with the
greatest accuracy by radio-occultation experiments. An important
consequence obtained is the fact that these planets have low
densities, from $0.688\gcc$ for Saturn to $1.64\gcc$ for Neptune, to
be compared with densities of 3.9 to 5.5$\gcc$ for the terrestrial
planets. Considering the compression that strongly increases with
mass, one is led naturally to the conclusion that these planets
contain an important proportion of light materials including hydrogen
and helium. It also implies that Uranus and Neptune which are less
massive must contain a relatively larger proportion of heavy elements
than Jupiter and Saturn. This may lead to a sub-classification between
the hydrogen-helium giant planets Jupiter and Saturn, and the ``ice
giants'' or ``sub giants'' Uranus and Neptune.

The planets are also relatively fast rotators, with periods of $\sim
10$ hours for Jupiter and Saturn, and $\sim 17$ hours for Uranus and
Neptune. The fact that this fast rotation visibly affects the figure
(shape) of these planets is seen by the significant difference between
the polar and equatorial radii. It also leads to gravitational moments
that differ significantly from a nul value. However, it is important
to stress that there is no unique rotation frame for these fluid
planets: atmospheric zonal winds imply that different latitude rotate
at different velocities (see \S~\ref{sec:dynamics}), and the magnetic
field provides an other rotation period. Because the latter is tied to
the deeper levels of the planet, it is believed to be more relevant
when interpreting the gravitational moments. The rotation periods
listed in Table~\ref{tab:moments} hence correspond to that of the
magnetic field. The case of Saturn appears to be complex and is
discussed in the next section.

\subsection{Magnetic fields}
\label{sec:magnetic fields}

As the Earth, the Sun and Mercury, our four giant planets possess
their own magnetic fields, as shown by the Voyager~2 measurements. The
structures of these magnetic fields are very different from one planet
to another and the dynamo mechanism that generates them is believed to
be related to convection in their interior but is otherwise essentially
unknown \citep[see][for a review]{1983RPPh...46..555S}.

The magnetic field $\bf B$ is generally expressed in form of a
development in spherical harmonics of the scalar potential $W$, such
that ${\bf B}=-\bfnabla W$:
\begin{equation}
W=a\sum_{n=1}^{\infty} \left(\frac{a}{r} \right)^{n+1}
\sum_{m=0}^n \left\{g_n^m \cos(m\phi)+h_n^m \sin(m\phi)\right\}
P_n^m(\cos\theta).
\label{eq:W}
\end{equation}
$r$ is the distance to the planet's center, $a$ its radius, $\theta$
the colatitude, $\phi$ the longitude and $P_n^m$ the associated
Legendre polynomials. The coefficients $g_n^m$ and $h_n^m$ are the
magnetic moments that characterize the field. They are expressed in
magnetic field units.

One can show that the first coefficients of relation~\eref{W} (for
$n=0$ and $n=1$) correspond to the potential of a magnetic dipole such
that $W={\bf M\cdot r}/r^3$ of moment:
\beq
M=a^3 \left\{\left(g_1^0\right)^2 + \left(g_1^1\right)^2 +
\left(h_1^1\right)^2\right\}^{1/2}.
\eeq

Jupiter and Saturn have magnetic fields of essentially dipolar nature,
of axis close to the rotation axis ($g_1^0$ is much larger than the
other harmonics); Uranus and Neptune have magnetic fields that are
intrinsically much more complex. To provide an idea of the intensity
of the magnetic fields, the value of the dipolar moments for the four
planets are $4.27\,\rm Gauss\,R_J^3$, $0.21\rm\,Gauss\,R_S^3$,
$0.23\rm\,Gauss\,R_U^3$, $0.133\rm\,Gauss\,R_N^3$, respectively
\citep{1982Natur.298...44C,1983JGR....88.8771A,1986Sci...233...85N,1989Sci...246.1473N}.

A true surprise from {\it Voyager} that has been confirmed by the {\it
Cassini-Huygens} mission is that Saturn's magnetic field is
axisymetric {\it to the limit of the measurement accuracy}: Saturn's
magnetic and rotation axes are perfectly aligned. Voyager measurements
indicated nevertheless a clear signature in the radio signal at 10h
39min 22s believed to be a consequence of the rotation of the magnetic
field. Determinations of a magnetic anomaly and new measurements by
Cassini have since considerably blurred the picture, and the
interpretation of the measurements have become unclear, with a deep
rotation period evaluated between that of Voyager and as slow as 10h
47min 6s \citep{2005Sci...307.1255G,2006Natur.441...62G} \citep[see
also][]{2000JGR...10513089G,2005JGRA..11012203C}. Note that models
discussed hereafter have not yet included this additional uncertainty.

\subsection{Atmospheric compositions}

In fluid planets, the distinction between the atmosphere and the
interior is not obvious. We name ``atmosphere'' the part of the planet
which can directly exchange radiation with the exterior
environment. This is also the part which is accessible by remote
sensing. It is important to note that the continuity between the
atmosphere and the interior does not guarantee that compositions
measured in the atmosphere can be extrapolated to the deep interior,
even in a fully convective environment: Processes such as phase
separations \citep[e.g.][]{Salpeter73, SS77a, FH03}, phase transitions
\citep[e.g.][]{1989oeps.book..539H}, chemical reactions
\citep[e.g.][]{FL94} can occur and decouple the surface
and interior compositions. Furthermore, imperfect mixing may also
occur, depending on the initial conditions
\citep[e.g.][]{1985Icar...62....4S}.

The conventional wisdom is however that these processes are limited to
certain species (e.g. helium) or that they have a relatively small
impact on the global abundances, so that the hydrogen-helium envelopes
may be considered relatively uniform, from the perspective of the
global abundance in heavy elements. We first discuss measurements 
made in the atmosphere before inferring interior compositions from
interior and evolution models. 

\subsubsection{Hydrogen and helium}

The most important components of the atmospheres of our giant planets
are also among the most difficult to detect: H$_2$ and He have a zero
dipolar moment and hence absorb very inefficiently visible and
infrared light. Their infrared absorption becomes important only at
high pressures when collision-induced absorption becomes significant
\citep[e.g.][]{1997AA...324..185B}. On the other hand, lines due to electronic transitions
correspond to very high altitudes in the atmosphere, and bear little
information on the structure of the deeper levels.  The only robust
result concerning the abundance of helium in a giant planet is by {\it
in situ} measurement by the Galileo probe in the atmosphere of Jupiter
\citep{1998JGR...10322815V}. The helium mole fraction (\ie\ number of
helium atoms over the total number of species in a given volume) is
$q_{\rm He}=0.1359\pm 0.0027$. The helium mass mixing ratio $Y$ (\ie\
mass of helium atoms over total mass) is constrained by its ratio over
hydrogen, $X$: $Y/(X+Y)=0.238\pm 0.05$. This ratio is by coincidence
that found in the Sun's atmosphere, but because of helium
sedimentation in the Sun's radiative zone, it was larger in the
protosolar nebula: $Y_{\rm proto}=0.275\pm 0.01$ and $(X+Y)_{\rm
proto}\approx 0.98$ \citep[e.g.][]{1995RvMP...67..781B}. 
Less helium is therefore found in the atmosphere
of Jupiter than inferred to be present when the planet formed. We will
discuss the consequences of this measurement later: let us mention
that the explanation invokes helium settling due to a phase separation
in the interiors of massive and cold giant planets.

Helium is also found to be depleted compared to the protosolar value
in Saturn's atmosphere. However, in this case the analysis is
complicated by the fact that Voyager radio occultations apparently led
to a wrong value. The current adopted value is now $Y=0.18-0.25$
\citep{CG00}, in agreement with values predicted by
interior and evolution models \citep{Guillot99b,Hubbard+99}. 
Finally, Uranus and Neptune are found to have near-protosolar helium
mixing ratios, but with considerable uncertainty. 

\begin{table}[htb]
\begin{center}
\parbox{12cm}{
\caption{Main gaseous components of heavy elements measured in the troposphere 
of giant planets}\label{tab:comp}
}
\scriptsize
\begin{tabular}{lllll} \hline\hline\vspace{.3ex}
 & {\bf Species} & {\bf Mixing ratio/H$_2$} & {\bf References} & {\bf Comments}
\vspace{.3ex}\\ \hline

{\bf Jupiter} & CH$_4$ & $(2.37 \pm 0.37)\times 10^{-3}$ &  \citet{Wong+04} & GPMS on Galileo(1) \\
& NH$_{3}$ & $(6.64 \pm 2.34) \times 10^{-3}$ & \citet{Wong+04} &
    {\it idem} \\
& H$_{2}$S & $(8.9 \pm 2.1) \times 10^{-3}$ & \citet{Wong+04} &
    {\it idem} \\
& H$_{2}$0 & $(4.9 \pm 1.6) \times 10^{-4}$ & \citet{Wong+04} &
    {\it idem}; region not well mixed \\
& $^{36}$Ar & $(6.1\pm 1.2) \times 10^{-6}$ & \citet{Atreya+99} & {\it idem} \\
& $^{84}$Kr & $(1.84\pm 0.37) \times 10^{-9}$ & \citet{Atreya+99} & {\it idem} \\
& $^{132}$Xe & $(4.9\pm 1.0) \times 10^{-11}$ & \citet{Atreya+99} & {\it
    idem} \medskip\\

%{\bf Saturn} & CH$_{4}$ & $(4.3\pm 1)\times 10^{-3}$ & \parbox{14em}{\citet{Flasar+05},\\revised by Orton et al. (2005)} & CIRS on Cassini (3) \\
{\bf Saturn} & CH$_{4}$ & $(4.3\pm 1)\times 10^{-3}$ & \citet{Flasar+05} & CIRS on Cassini (3) \\
 & NH$_{3}$ & $(1\pm 1)\times 10^{-4}$ & \citet{BS89} &
Ground-based microwave (4) \\
 & H$_{2}$S & $(2.2\pm 0.3)\times 10^{-4}$ & \citet{BS89} &
{\it idem} \medskip\\

{\bf Uranus} & CH$_{4}$ & $(3.3\pm 1.1)\times 10^{-2}$ & \citet{Gautier+95} & \parbox{14em}{Compilation from ground-based observations} \\
 & H$_{2}$S & $(1\pm 1)\times 10^{-4}$ & \citet{BS89} &
Ground-based microwave (4) \medskip\\

{\bf Neptune} & CH$_{4}$ & $(3.3\pm 1.1)\times 10^{-2}$ & \citet{Gautier+95} & \parbox{14em}{Compilation from ground-based observations} \\
 & H$_{2}$S & $(7.5\pm 3.25)\times 10^{-4}$ & \citet{dPRA91} &
Ground based microwave (5) \\
& H$_{2}$O & $7.7\times 10(-1)$ & \citet{LF94} & Inferred
from CO (6)\\ \hline
\end{tabular}
\end{center}
\noindent
(1) Galileo Probe Mass Spectrometer aboard the atmospheric probe 
in Jupiter\\
(2) The signal stopped at the 22 bar levels prior to have reached 
a constant value. It is currently believed that the region where 
the probe made measurementds was atypically dry and that the 
bulk abundance of H$_{2}$O in Jupiter has not been measured.\\
(3) Composite Infra Red Spectrometer aboard the Cassini spacecraft\\
(4) Ground based measurements of the microwave continuum. The 
result is somewhat uncertain due to the difficulty to precisely 
estimate opacities of absorbing species\\
(5) Ground based microwave measurements. An oversolar H$_{2}$S abundance 
is required to interpret the depletion of NH$_{3}$ in the upper 
troposphere\\
(6) Inferred from the microwave detection of CO in the troposphere. 
Note that the validity of the approach is questionned by
\citet{Bezard+02}. A large amount of water seems to be present anyway
in the deep atmosphere of Neptune. The case of Uranus is still uncertain 
because it is not known so far if CO is present in the troposphere 
of the planet.
\end{table}

\subsubsection{Heavy elements}

The abundance of other elements than hydrogen and helium (that we will
call hereafter ``heavy elements'') bears crucial information for the
understanding of the processes that led to the formation of these
planets. 

The most abundant heavy elements in the envelopes of our four giant
planets are O, C, N, S. It is possible to model the chemistry of gases
in the tropospheres from the top of the convective zone down to the
2000 K temperature level \citep{FL94}. Models conclude
that, whatever the initial composition in these elements of
planetesimals which collapsed with hydrogen onto Jupiter and Saturn
cores during the last phase of the planetary formation, C in the upper
tropospheres of giant planets is mainly in the form of gaseous CH$_4$,
N in the form of NH$_3$, S in the form of H$_2$S, and O in the form of
H$_2$O. All these gases, but methane in Jupiter and Saturn, condense
in the upper troposphere, but vaporize at deeper levels when the
temperature increases. Interestingly enough, noble gases are not
expected to condense even at the cold tropopause temperatures of
Uranus and Neptune.

The mass spectrometer aboard the Galileo atmospheric probe has
performed in situ measurements of Ar, Kr, Xe, CH$_4$, NH$_3$, H$_2$S,
and H$_2$O in the troposphere of Jupiter. C, N, and S were found to be
oversolar by a factor 3 to 4 \citep{Wong+04}, which was not
unexpected because condensation of nebula gases results in enriching
icy grains and planetesimals. The surprise came from Ar, Kr, Xe, which
were expected to be solar because they are difficult to condense, but
turned out to be oversolar by factors 2 to 4 \citep{Owen+99,Wong+04}. One exception among these enriched species was neon,
which was found to be significantly undersolar, but was predicted to
be so because of a capture by the falling helium droplets
\citep{RS95}. Another exception was water, but this molecule is 
affected by meteorological processes, and the probe was shown to have
fallen into a dry region of Jupiter's atmosphere.

Specifically, CH$_4$/H$_2$ has been found oversolar in the four giant
planets: the C/H ratio corresponding to the measured abundances
is always higher than the solar C/H ratio, and in fact appears to be
increasing with distance to the Sun. C/H is 3, 7.5, 45 and 45 times
solar, in Jupiter, Saturn, Uranus and Neptune, respectively. Note
that the quoted enrichments are subject to changes when the solar
abundances tables are revised, which happens surprisingly frequently. 

Except for Jupiter, the determination of the NH$_3$ abundance is more
uncertain than that of CH$_4$ because it is model dependant. It is
derived from fitting microwave spectra of giant planets which exhibit
a continuum opacity, more difficult to model than absorption spectral
lines. However, the N/H enrichment seems to be, so far, fairly
constant from a planet to another, around a factor 2, so that C/N is
higher in Saturn than in Jupiter, and still higher in Uranus and
Neptune. 

H$_2$S has been measured in situ in Jupiter, but in the three other giant
planets its large abundance is derived from the requirement to deplete
NH3 at deeper levels than the saturation one; This scenario has been
proposed a long time ago by \citet{GJO78}. It implies that S/H
is substantially oversolar in Uranus and in Neptune.

H$_2$O is difficult to measure in all four giant planets because of
its condensation relatively deep. It was hoped that the Galileo probe
would provide a measurement of its deep abundance, but the probe fell
into one of Jupiter's 5-microns hot spot, what is now believed to be a
dry region mostly governed by downwelling motions
\citep[e.g.][]{SI98}. As a result, and although the probe provided
measurements down to 22 bars, well below water's canonical 5 bar cloud
base, it is believed that this measurement of a water abundance equal
to a fraction of the solar value is only a lower limit.

\subsection{Energy balance and atmospheric temperature profiles}

Jupiter, Saturn and Neptune are observed to emit significantly more
energy than they receive from the Sun (see Table~\ref{tab:flux}). The
case of Uranus is less clear. Its intrinsic heat flux $F_{\rm int}$ is
significantly smaller than that of the other giant planets. Detailed
modeling of its atmosphere however indicate that $F_{\rm
int}\wig{>}60\rm\,erg\,cm^{-2}\,s^{-1}$ \citep{MM99}.  With
this caveat, all four giant planets can be said to emit more energy
than they receive from the Sun.  \citet{Hubbard68} showed in the case of
Jupiter that this can be explained simply by the progressive
contraction and cooling of the planets.
% Contrary to
% what is sometimes found in the litterature, there is {\it no} lacking source
% of energy required to explain the giant planets' energy budget.
% However, we will see that significant uncertainties arise from our
% limited knowledge of heat transport in the atmosphere and in the
% interior.

\begin{table}[htb]
\begin{center}
\caption{Energy balance as determined from Voyager IRIS data\ti{a}.}
\label{tab:flux}
\small
\begin{tabular}{l r@{$\pm$}l r@{$\pm$}l r@{$\pm$}l r@{$\pm$}l} \hline \hline
	& \multicolumn{2}{c}{\bf Jupiter} &
\multicolumn{2}{c}{\bf Saturn} &\multicolumn{2}{c}{\bf Uranus} 
&\multicolumn{2}{c}{\bf Neptune}  \\
\hline 
Absorbed power [$10^{16}$\, J\,s$^{-1}$] & 50.14&2.48 &
11.14&0.50 & 0.526&0.037 & 0.204&0.019 \\
Emitted power [$10^{16}$\, J\,s$^{-1}$] & 83.65&0.84 &
19.77&0.32 & 0.560&0.011 & 0.534&0.029 \\
Intrinsic power [$10^{16}$\, J\,s$^{-1}$]&
33.5& 2.6 & 8.63&0.60 & 
\multicolumn{2}{c}{0.034\,\parbox{2em}{\tiny $+0.038$ $-0.034$}} 
& 0.330& 0.035 \\
%L/M$\,^{\rm (c)}$ [$10^7$\, erg.g$^{-1}$.s$^{-1}$] & $17.6& 1.4$ &
%$15.2& 1.1$ & 0.392\,\parbox{2em}{\tiny $+0.441$ $-0.392$} & 
%$3.22& 0.34$ \\
Intrinsic flux [J\,s$^{-1}$\,m$^{-2}$] & 5.44&
0.43 & 2.01& 0.14 & \multicolumn{2}{c}{0.042\,\parbox{1.2em}{\tiny
$+0.047$ $-0.042$}} & 0.433& 0.046 \\
Bond albedo [] & 0.343&0.032 & 0.342&0.030 & 0.300&0.049 & 0.290&0.067 \\
Effective temperature [K] & 124.4& 0.3 & 95.0&
0.4 & 59.1& 0.3 & 59.3& 0.8 \\ 
1-bar temperature\ti{b} [K] & 165&5 & 135&5 & 76&2 & 72&2 \\
\hline\hline 
\multicolumn{9}{l}{\ti{a} After \citet{PC91}} \\
\multicolumn{9}{l}{\ti{b} \citet{Lindal92}} 
\end{tabular}
\normalsize
\end{center}
\end{table}

A crucial consequence of the presence of an intrinsic heat flux is
that it requires high internal temperatures ($\sim 10,000\K$ or more),
and that consequently the giant planets are {\it fluid} (not solid)
(\cite{Hubbard68}; see also \cite{Hubbard+95}). Another consequence is
that they are essentially
convective, and that their interior temperature profile are close to
{\it adiabats}. We will come back to this in more detail. 

\begin{figure}[htb]
\resizebox{12cm}{!}{\includegraphics{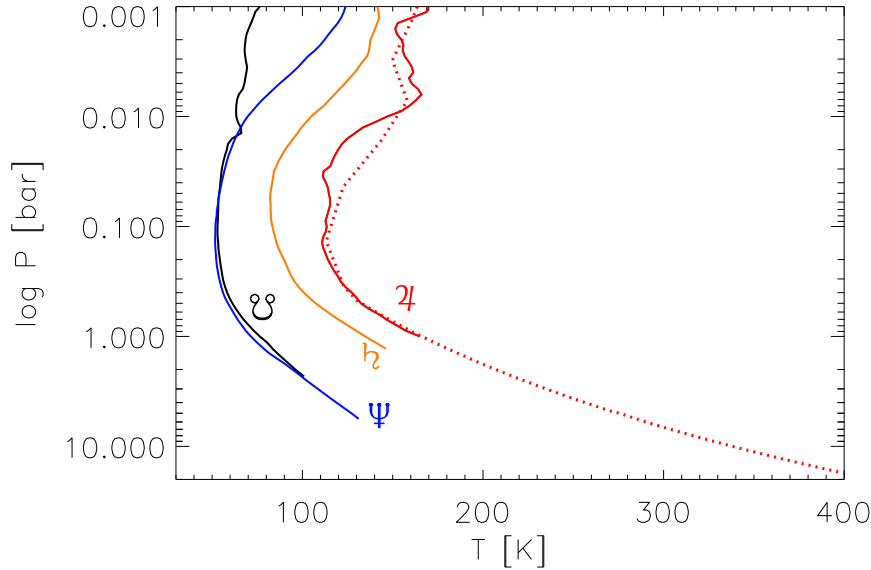}}
\caption{Atmospheric temperatures as a function of pressure for
Jupiter, Saturn, Uranus and Neptune, as obtained from Voyager
radio-occultation experiments \citep[see][]{Lindal92}. The dotted line
corresponds to the temperature profile retrieved by the Galileo probe,
down to 22\bar\ and a temperature of $428\K$ \citep{Seiff+98}.}
\label{fig:atm_temp}
\end{figure}

The deep atmospheres (more accurately tropospheres) of the four giant
planets are indeed observed to be close to adiabats, a result first
obtained by spectroscopic models \citep{Trafton67}, then verified by
radio-occultation experiments by the Voyager spacecrafts, and by the
{\it in situ} measurement from the Galileo probe
(fig.~\ref{fig:atm_temp}). The temperature profiles show a temperature
minimum, in a region near 0.2\bar\, called the tropopause. At higher
altitudes, in the stratosphere, the temperature gradient is negative
(increasing with decreasing pressure). In the regions that we will be
mostly concerned with, in the troposphere and in the deeper interior,
the temperature always increases with depth. It can be noticed that
the slope of the temperature profile in fig~\ref{fig:atm_temp} becomes
almost constant when the atmosphere becomes convective, at pressures
of a few tens of bars, in the four giant planets.

It should be noted that the 1 bar temperatures listed in
table~\ref{tab:flux} and the profiles shown in Fig.~\ref{fig:atm_temp}
are retrieved from radio-occultation measurements using a helium to
hydrogen ratio which, at least in the case of Jupiter and Saturn, was
shown to be incorrect. The new values of $Y$ are found to lead to
increased temperatures by $\sim 5\K$ in Jupiter and $\sim 10\K$ in
Saturn \citep[see][]{Guillot99a}. However, the Galileo probe found a 1 bar
temperature of $166\K$ \citep{Seiff+98}, and generally a good
agreement with the Voyager radio-occultation profile with the wrong
He/H$_2$ value.

When studied at low spatial resolution, it is found that all four
giant planets, in spite of their inhomogeneous appearances, have a
rather uniform brightness temperature, with pole-to-equator
latitudinal variations limited to a few kelvins
\citep[\eg][]{Ingersoll+95}. However, in the case of Jupiter, some
small regions are known to be very different from the average of the
planet. This is the case of hot spots, which cover about 1\% of the
surface of the planet at any given time, but contribute to most of the
emitted flux at 5 microns, due to their dryness (absence of water
vapor) and their temperature brightness which can, at this wavelength,
peak to 260\K.

\subsection{Atmospheric dynamics: winds and weather}
\label{sec:dynamics}
The atmospheres of all giant planets are evidently complex and
turbulent in nature. This can for example be seen from the mean zonal
winds (inferred from cloud tracking), which are very rapidly varying
functions of the latitude \citep[see \eg][]{Ingersoll+95}: while some
of the regions rotate at the same speed as the interior magnetic field
(in the so-called ``system III'' reference frame), most of the
atmospheres do not. Jupiter and Saturn both have superrotating
equators ($+100$ and $+400\m\sec^{-1}$ in system III, for Jupiter and
Saturn, respectively), Uranus and Neptune have subrotating equators,
and superrotating high latitude jets. Neptune, which receives the
smallest amount of energy from the Sun has the largest peak-to-peak
latitudinal variations in wind velocity: about $600\m\sec^{-1}$.  It
can be noted that, contrary to the case of the strongly irradiated
planets to be discussed later, the winds of Jupiter, Saturn, Uranus
and Neptune, are significantly slower than the planet itself under its
own spin (from 12.2\,km$\sec^{-1}$ for Jupiter to 2.6\,km$\sec^{-1}$
for Neptune, at the equator).

The observed surface winds are believed to be related to motions in
the planets' interiors, which, according to the Taylor-Proudman
theorem, should be confined by the rapid rotation to the plane
perpendicular to the axis of rotation
\citep[\eg][]{Busse78}. Unfortunately, no convincing model is yet
capable of modeling with sufficient accuracy both the interior and the
surface layers.

Our giant planets also exhibit planetary-scale to small-scale storms
with very different temporal variations. For example, Jupiter's great
red spot is a 12000\,km-diameter anticyclone found to have lasted for
at least 300 years \citep[e.g.][]{SimonMiller+02}. Storms developing
over the entire planet have even been observed on Saturn
\citep{SanchezLavega+96}. Uranus and Neptune's storm system has been
shown to have been significantly altered since the Voyager era
\citep{Rages+02,Hammel+05}. On Jupiter, small-scale storms related to
cumulus-type cloud systems have been observed
\citep[e.g.][]{Gierasch+00,HSG02}, and lightning strikes have been
monitored by Galileo \citep[e.g.][]{Little+99}. These represent only a
small arbitrary subset of the work concerning the complex atmospheres
of these planets.

It is tempting to extrapolate these observations to the objects
outside our Solar System as well. However, it is important to stress
that an important component of the variability in the atmospheres of
our giant planets is the presence of relatively abundant condensing
chemical species: ammonia and water in the case of Jupiter and Saturn,
and methane for Uranus and Neptune. These species can only condense
in very cold atmospheres, thus providing latent heat to fuel important
storms. Depending on their temperatures and compositions, extrasolar
planets may or may not possess such important condensing species
\citep[e.g.][]{Guillot99b}.

\subsection{Moons and rings}

A discussion of our giant planets motivated by the opportunity to
extrapolate the results to objects outside our solar system would be
incomplete without mentioning the moons and rings that these planets
all possess (see chapters by Breuer \& Moore, by Peale and by Husmann
et al.). First, the satellites/moons can be distinguished from their
orbital characteristics as regular or irregular. The first ones have
generally circular, prograde orbits. The latter tend to have
eccentric, extended, and/or retrograde orbits. 

These satellites are numerous: After the Voyager era, Jupiter was
known to possess 16 satellites, Saturn to have 18, Uranus 20 and
Neptune 8.  Recent extensive observation programs have seen the number
of satellites increase considerably, with a growing list of satellites
presently reaching 62, 56, 27 and 13 for Jupiter, Saturn, Uranus and
Neptune, respectively. All of the new satellites discovered since
Voyager are classified as irregular.

The presence of regular and irregular satellites is due in part to the 
history of planet formation. It is believed that the regular
satellites have mostly been formed in the protoplanetary
subnebulae that surrounded the giant planets (at least Jupiter and
Saturn) at the time when they accreted their envelopes. On the other
hand, the irregular satellites are thought to have been captured by
the planet. This is for example believed to be the case of Neptune's
largest moon, Triton, which has a retrograde orbit. 

A few satellites stand out by having relatively large masses: it is
the case of Jupiter's Io, Europa, Ganymede and Callisto, of Saturn's
Titan, and of Neptune's Triton. Ganymede is the most massive of them,
being about twice the mass of our Moon. However, compared to the mass
of the central planet, these moons and satellites have very small
weights: $10^{-4}$ and less for Jupiter, $1/4000$ for Saturn,
$1/25000$ for Uranus and $1/4500$ for Neptune. All these satellites
orbit relatively closely to their planets. The farthest one,
Callisto revolves around Jupiter in about 16 Earth days. 

The four giant planets also have rings, whose material is probably 
constantly resupplied from their satellites. The ring of Saturn stands
out as the only one directly visible with binoculars. In this
particular case, its enormous area allows it to reflect a sizable
fraction of the stellar flux arriving at Saturn, and makes this
particular ring as bright as the planet itself. The occurrence of such
rings would make the detection of extrasolar planets slightly easier,
but it is yet unclear how frequent they can be, and how close to the
stars rings can survive both the increased radiation and tidal
forces.

%% \subsection{Oscillations}

%% Last but not least, the case for the existence of free oscillations of
%% the giant planets is still unresolved. Such a discovery would lead to
%% great leaps in our knowledge of the interior of these planets, as can
%% be seen from the level of accuracy reached by solar interior models
%% since the discovery of its oscillations. Observations aimed at
%% detecting modes of Jupiter have shown promising results (Schmider et
%% al. 1991), but have thus far been limited by instrumental and
%% windowing effects. A recent work by Mosser \etal\ (2000) puts an upper
%% limit to the amplitude of the modes at $0.6\m\sec^{-1}$, and shows an
%% increased energy of the Fourier spectrum in the expected range of
%% frequencies. Observations from space of through an Earth-based network
%% should be pursued in order to verify these results.

\subsection{Extrasolar planets}

\begin{figure}[htb]
\resizebox{12cm}{!}{\includegraphics{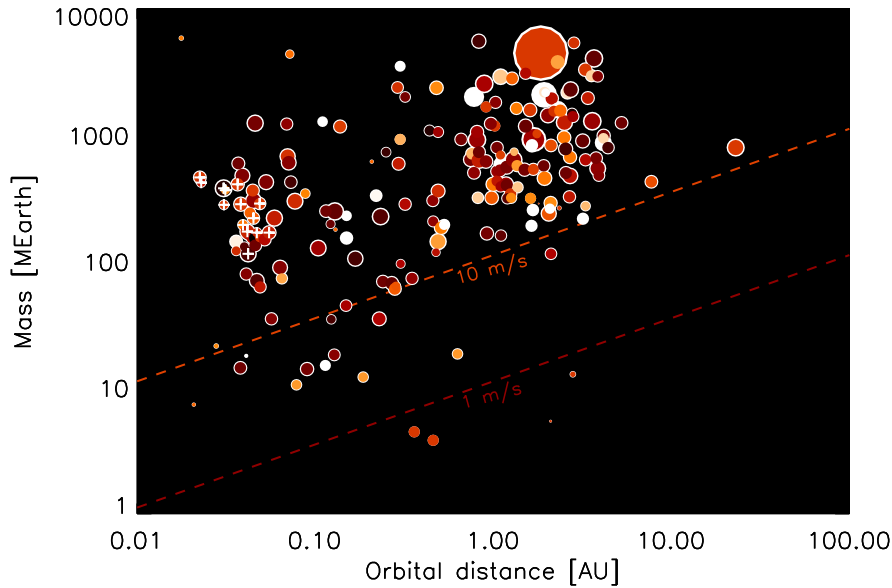}}
\caption{Masses and orbital distances of the extrasolar planets
  discovered by 2006. The size of the symbols is proportional to the
  mass of the parent star (from 0.1 to 4 stellar masses). The color
  (from white to red and black) is proportional to the stellar
  metallicity. Stars with metallicities $\rm[Fe/H]<0.2$ are shown in
  white. The radial velocimetry thresholds at 1 and 10 m/s,
  encompassing the detection limits of most current surveys, are
  indicated as dashed lines. {\em Transiting} planets are highlighted
  with white crosses.}
\label{fig:exo}
\end{figure}

Huge progresses have been made in the field of extrasolar planets
since the detection of the first giant planet orbiting a solar-type
star by \citet{MQ95}. More than 200 planets are known at the
time of this review, and importantly, 14 planets that transit their
star at each orbital revolution have been identified (see
fig~\ref{fig:exo}). These transiting planets are especially
interesting because of the possibility to measure both their mass and
size and thus obtain constraints on their global composition.

In spite of their particular location just a few stellar radii away
from their stars, the transiting planets that have been discovered bear
some resemblance with their Solar System cousins in the sense that
they are also mostly made of hydrogen and helium
\citep[e.g.][]{Burrows+00, Guillot05, Baraffe+05}. They are, however,
much hotter due to the intense irradiation that they receive. 

Although obtaining direct informations on these planets represent a
great observational challenge, several key steps have been
accomplished: Atomic Sodium, predicted to be detectable \citep{SS00},
has indeed been detected by transit spectroscopy around
one planet \citep{CBNG02}. Hydrodynamically escaping
hydrogen, oxygen and carbon have also been detected from the same
planet \citep{VidalMadjar+03,VidalMadjar+04}. The measurement of the {\it
secondary} eclipse of several planets by the Spitzer satellite allowed
a constraint on spectral and hence thermal properties of the planetary
atmospheres \citep{Charbonneau+05,DSRH05}. Recently,
the light curve of a non-transiting planet was detected in the
infrared, also with Spitzer, providing preliminary indications of a
strong day/night temperature variation \citep{Harrington+06}
perhaps even larger than predicted \citep[see][]{SG02}.

Obviously, there is a big potential for growth in this young field,
and the comparison between fine observations made for giant planets in
our Solar System and the more crude, but also more statistically
significant data obtained for planets around other stars promises to
be extremely fruitful to better understand these objects.

\section{The calculation of interior and evolution models}

\subsection{Basic equations}\label{sec:basic}

The structure and evolution of a
giant planet is governed by the following hydrostatic, thermodynamic,
mass conservation and energy conservation equations:
\begin{eqnarray}
\dpar{P}{r}&=&-\rho g \label{eq:dpdr}\\
{\partial T\over\partial r}&=&{\partial P\over \partial r}{T\over
P}\nabla_T. \label{eq:dtdr}\\
{\partial m\over\partial r}&=&4\pi r^2\rho. \label{eq:dmdr}\\
{\partial L\over\partial r}&=&4\pi r^2\rho \left(\dot{\epsilon}-
T{\partial S\over \partial t}\right),\label{eq:dldr}
\end{eqnarray}
where $P$ is the pressure, $\rho$ the density, and $g=Gm/r^2$ the
gravity ($m$ is the mass, $r$ the radius and $G$ the gravitational
constant). The temperature gradient $\nabla_T\equiv(d\ln T/d\ln P)$
depends on the process by which the internal heat is transported.
$L$ is the intrinsic luminosity, $t$ the time, $S$ the
specific entropy (per unit mass), and $\dot{\epsilon}$ accounts for
the sources of energy due e.g. to radioactivity or more importantly
nuclear reactions. Generally it is a good approximation to assume
$\dot{\epsilon}\sim 0$ for objects less massive than $\sim 13\mjup$,
i.e. too cold to even burn deuterium (but we will see that in certain
conditions this term may be useful, even for low mass planets). 

The boundary condition at the center is trivial: $r=0$; ($m=0$,
$L=0$). The external boundary condition is more difficult to obtain
because it depends on how energy is transported in the atmosphere. One
possibility is to use the Eddington approximation, and to write
\citep[e.g.][]{Chandrasekhar39}: $r=R$; ($T_0=\teff$,
$P_0=2/3\,g/\kappa$), where $\teff$ is the effective temperature
(defined by $L=4\pi R\sigma\teff^4$, with $\sigma$ being the
Stephan-Boltzmann constant), and $\kappa$ is the opacity. Note for
example that in the case of Jupiter $\teff=124$\,K,
$g=26\rm\,m\,s^{-2}$ and $\kappa\approx 5\times 10^{-3}
(P/1\rm\,bar)\,m^2\,kg^{-1}$. This implies $P_0\approx 0.2$\,bar
(20,000\,Pa), which is actually close to Jupiter's tropopause, where
$T\approx 110$\,K.

More generally, one has to use an atmospheric model relating the
temperature and pressure at a given level to the radius $R$, intrinsic
luminosity $L$ and incoming stellar luminosity $\linc$: $r=R$;
($T_0=T_0(R,L,\linc)$, $P_0=P_0(R,L,\linc)$). $P_0$ is chosen to
satisfy the condition that the corresponding optical depth at that
level should be much larger than unity. If the stellar flux is
absorbed mostly in a convective zone, then the problem can be
simplified by using $T_0(R,L,\linc)\approx T_0(R,L+\linc,0)$
\citep[e.g.][]{Hubbard77}. An example of such a model is described by
\citet{Saumon+96} and \citet{HBL02} and is used hereafter to model the
planets in the low irradiation limit.

\subsection{High pressure physics \& equations of state}

\begin{figure}[htbp]
\hspace*{-2cm}
\resizebox{16cm}{!}{\includegraphics[angle=90]{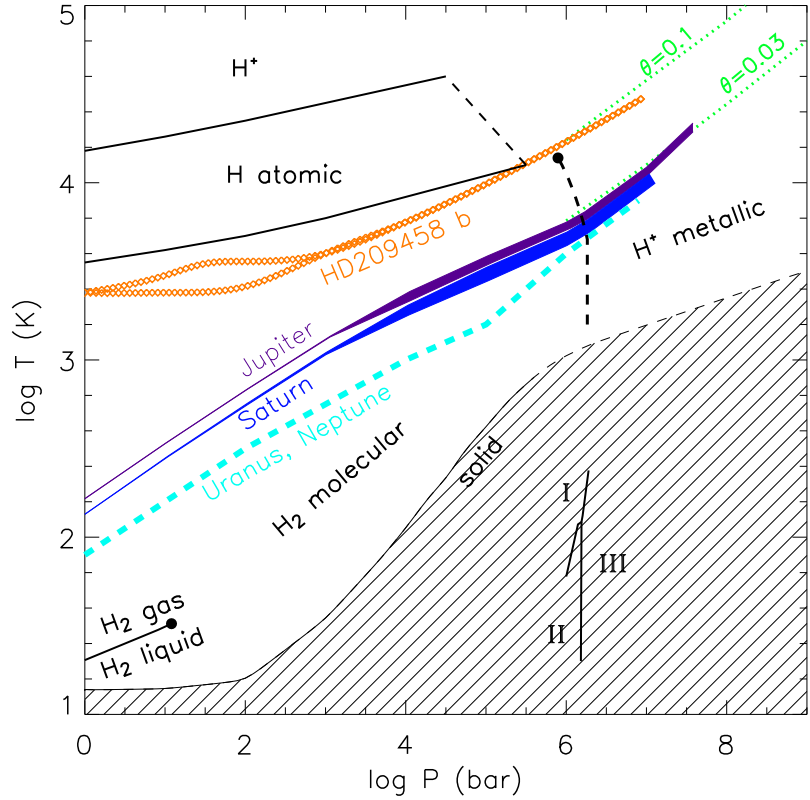}}
%\centerline{\hbox{\psfig{file=diag_phase.ps,width=14cm}}}
\caption{Phase diagram for hydrogen with the main phase transitions
occurring in the fluid or gas phase. The temperature-pressure profiles
for Jupiter, Saturn, Uranus, Neptune, and the exoplanet HD\,209458\,b are
shown. The dashed nearly vertical line near 1\,Mbar is indicative of
the molecular to metallic transition (here it represents the so-called
plasma phase transition as calculated by \citet{SCvH95}). The
region in which hydrogen is in solid phase \citep{DLL00,Gregoryanz+03}
is represented as a hatched area. The three 
phases (I,II,III) of solid hydrogen are shown
\citep[see][]{MH94}. Values of the degeneracy parameter $\theta$ are
indicated as 
dotted lines to the upper right corner of the figure.}
\label{fig:phase_diag}
\end{figure}

In terms of pressures and temperatures, the interiors of giant planets
lie in a region for which accurate equations of state (EOS) are
extremely difficult to calculate. This is because both molecules,
atoms, and ions can all coexist, in a fluid that is partially degenerate
(free electrons have energies that are determined both by quantum and
thermal effects) and partially coupled (coulomb interactions
between ions are not dominant but must be taken into account). The
presence of many elements and their possible interactions further
complicate matters. For lack of space, this section will mostly focus
on hydrogen whose EOS has seen the most important developments in
recent years. A phase diagram of hydrogen (fig.~\ref{fig:phase_diag})
illustrates some of the important phenomena that occur in giant
planets.

The photospheres of giant planets are generally relatively cold (50 to
3000\,K) and at low pressure (0.1 to 10\,bar, or $10^4$ to
$10^6$\,Pa), so that hydrogen is in molecular form and the perfect gas
conditions apply.  As one goes deeper into the interior hydrogen and
helium progressively become fluid. (The perfect gas relation tends to
underestimate the pressure by 10\% or more when the density becomes
larger than about $0.02\gcc$ ($P\wig{>} 1\,$kbar in the case of
Jupiter)).

Characteristic interior pressures are considerably larger however: as
implied by Eqs.~\ref{eq:dpdr} and \ref{eq:dmdr}, $P_{\rm c}\approx
GM^2/R^4$, of the order of 10-100\,Mbar for Jupiter and Saturn. At
these pressures and the corresponding densities, the Fermi temperature
$T_{\rm F}$ is larger than $10^5$\,K. This implies that electrons are
degenerate. Figure~\ref{fig:phase_diag} shows that inside Jupiter,
Saturn, the extrasolar planet HD\,209458\,b, but also for giant
planets in general for most of their history, the degeneracy parameter
$\theta=T/T_{\rm F}$ is between 0.1 and 0.03. Therefore, the energy of
electrons in the interior is expected to be only slightly larger than
their non-relativistic, fully degenerate limit: $u_{\rm e}\ge
3/5\,kT_{\rm F} =15.6\left(\rho/\mu_{\rm e}\right)^{2/3}\ \rm eV$,
where $k$ is Boltzmann's constant, $\mu_{\rm e}$ is the number of
electrons per nucleon and $\rho$ is the density in $\rm
g\,cm^{-3}$. For pure hydrogen, when the density reaches $\sim
0.8\gcc$, the average energy of electrons becomes larger than
hydrogen's ionization potential, even at zero temperature: hydrogen
pressure-ionizes and becomes metallic. This molecular to metallic
transition occurs near Mbar pressures, but exactly how this happens
remains unclear because of the complex interplay of thermal,
coulomb, and degeneracy effects (in particular, whether hydrogen
metallizes into an atomic state H$^+$ --- as suggested in
Fig.~\ref{fig:phase_diag} --- or first metallizes in the molecular
state H$_2$ remains to be clarified).

Recent laboratory measurements on fluid deuterium have been able to
reach pressures above $\wig> 1\,$Mbar, and provide new data in a
region where the EOS remains most uncertain. Gas-guns experiments have
been able to measure the reshock temperature \citep{HRN95}, near
$T\sim 5000\K$, $P\sim 0.8\,$Mbar, and a rise in the conductivity of
molecular hydrogen up to $T\sim 3000\K$, $P\sim 1.4\,$Mbar, a sign
that metallicity may have been reached \citep{WMN96}. The following
few years have seen the development of laser-induced shock compression
\citep{daSilva+97,Collins+98}, pulsed-power shock compression
\citep{Knudson+04}, and convergent shock wave experiments
\citep{Belov+02, Boriskov+05} in a high-pressure ($P=0.3-4\,$Mbar)
high-temperature ($T\sim 6000-10^5\K$) regime. Unfortunately,
experimental results along the principal Hugoniot of deuterium do not
agree in this pressure range.  Laser compression data give a maximum
compression of $\sim 6\,$ while both the pulsed-power compression
experiments and the convergent shock wave experiments find a value of
$\sim 4\,$. Models that are partly calibrated with experimental data
\citep{SCvH95,Ross98,RY01} obtain a generally good agreement with the
laser-compression data. However, the fact that independant models
based on first principles \citep{Militzer+01,Desjarlais03,BMG04} yield
low compressions strongly favors this solution.

The question of the existence of a first-order molecular to metallic
transition of hydrogen (i.e. both molecular dissociation and
ionisation occur simultaneously and discontinuously at the so-called
plasma phase transition, or PPT) remains however. The critical line
shown in \rfig{phase_diag} corresponds to calculations by
\citet{SCvH95}, but may be caused by artefacts in the free energy 
calculation. Recent Density Functional Theory (DFT) simulations by
\citet{BMG04} indicate the possibility of a first order
liquid-liquid transition but other path-integral calculations
\citep{Militzer+01} do not. It is crucial to assess the
existence of such a PPT because it would affect both convection and
chemical composition in the giant planets.

A clear result from \rfig{phase_diag} at least is that, as first shown
by \citet{Hubbard68}, the interiors of the hydrogen-helium giant planets
are {\it fluid}, whatever their age: of course, they avoid the
critical point for the liquid gas transition in hydrogen and helium,
at very low temperatures, but they also lie comfortably above the
solidification lines for hydrogen and helium. (An {\it isolated}
Jupiter should begin partial solidification only after at least $\sim
10^3\,$Ga of evolution.) They are considered to be fluid because at
the high pressures and relatively modest temperatures in their
interiors, coulomb interactions between ions play an important role in
the EOS and yield a behavior that is more reminiscent of that of a
liquid than that of a gas, contrary to what is the case in
e.g. solar-like stars.  For Uranus and Neptune, the situation is
actually more complex because at large pressures they are not expected
to contain hydrogen, but numerical simulations show that ices in their
interior should be fluid as well \citep{Cavazzoni+99}.

Models of the interiors of giant planets require thermodynamically
consistent EOSs calculated over the entire domain of pressure and
temperature spanned by the planets during their evolution. Elements
other than hydrogen, most importantly helium, should be consistently
included. Such a calculation is a daunting task, and the only recent
attempt at such an astrophysical EOS for substellar objects is that by
\citet{SCvH95}. Another set of EOSs reproducing either the high-
or low-compression results was calculated by \citet{SG04}
specifically for the calculation of present-day models of Jupiter and
Saturn.

These EOSs have so far included other elements (including helium),
only in a very approximative way, i.e. with EOSs for helium and heavy
elements that are based on interpolations between somewhat ideal
regimes, using an additive volume law, and neglecting the possibility
of existence of phase separations \citep[see][for further
  discussions]{HBL02, GSHS04}.

\subsection{Heat transport}

Giant planets possess hot interiors, implying that a
relatively large amount of energy has to be transported from the deep
regions of the planets to their surface. This can either be done by
radiation, conduction, or, if these processes are not sufficient, by
convection. Convection is generally ensured by the rapid rise of the
opacity with increasing pressure and temperature.  At pressures of a
bar or more and relatively low temperatures (less than 1000\,K), the
three dominant sources of opacities are water, methane and
collision-induced absorption by hydrogen molecules.

However, in the intermediate temperature range between $\sim 1200$ and
$1500\K$, the Rosseland opacity due to the hydrogen and helium
absorption behaves differently: the absorption at any given wavelength
increases with density, but because the temperature also rises, the
photons are emitted at shorter wavelengths, where the monochromatic
absorption is smaller. As a consequence, the opacity can
decrease. This was shown by Guillot et al. (1994) to potentially lead
to the presence of a deep radiative zone in the interiors of Jupiter,
Saturn and Uranus.

This problem must however be reanalyzed in the light of recent
observations and analyses of brown dwarfs. Their spectra show
unexpectedly wide sodium and potassium absorption lines (see Burrows,
Marley \& Sharp 2000), in spectral regions where hydrogen, helium,
water, methane and ammonia are relatively transparent. It thus appears
that the added contribution of these elements (if they are indeed
present) would wipe out any radiative region at these
levels \citep{GSHS04}. 

At temperatures above $1500\sim 2000\K$ two important sources of
opacity appear: (i) the rising number of electrons greatly enhances
the absorption of H$_2^-$ and H$^-$; (ii) TiO, a very strong absorber
at visible wavelengths is freed by the vaporization of
CaTiO$_3$. Again, the opacity rises rapidly which ensures a convective
transport of the heat. Still deeper, conduction by free electrons
becomes more efficient, but the densities are found not to be high
enough for this process to be significant, except perhaps near the
central core \citep[see][]{Hubbard68, SS77a}.

While our giant planets seem to possess globally convective interiors,
strongly irradiated extrasolar planets must develop a radiative zone
just beneath the levels where most of the stellar irradiation is
absorbed. Depending on the irradiation and characteristics of the
planet, this zone may extend down to kbar levels, the deeper levels
being convective. In this case, a careful determination of the
opacities is necessary (but generally not possible) as these control
the cooling and contraction of the deeper interior \citep[see][for a discussion of opacities and tables for substellar
atmospheres and interiors]{Ferguson+05}.

\subsection{The contraction and cooling histories of giant planets}

The interiors of giant planets is expected to evolve with time from a
high entropy, high $\theta$ value, hot initial state to a low entropy,
low $\theta$, cold degenerate state. The essential physics behind can
be derived from the well-known virial theorem and the energy
conservation which link the planet's internal energy $E_{\rm i}$,
gravitational energy $E_{\rm g}$ and luminosity through:
\begin{eqnarray}
\xi E_{\rm i} + E_{\rm g} &=&0,\\
L &=& -{\xi-1\over \xi}{dE_{\rm g}\over dt},
\end{eqnarray}
where $\xi=\int_0^M 3(P/\rho)dm / \int_0^M u dm\approx <\!3P/\rho
u\!>$, the brackets indicating averaging, and $u$ is the specific
internal energy. For a diatomic perfect gas, $\xi=3.2$; for
fully-degenerate non-relativistic electrons, $\xi=2$.

Thus, for a giant planet or brown dwarf beginning its life mostly as a
perfect H$_2$ gas, two third of the energy gained by contraction is
radiated away, one third being used to increase $E_{\rm i}$. The
internal energy being proportional to the temperature, the effect is
to heat up the planet. This represents the slightly counter-intuitive
but well known effect that a star or giant planet initially heats up
while radiating a significant luminosity \citep[e.g.][]{KW94}.

Let us now move further in the evolution, when the contraction has
proceeded to a point where the electrons have become degenerate.  For
simplicity, we will ignore coulomb interactions and exchange terms,
and assume that the internal energy can be written as $E_{\rm
i}=E_{\rm el}+E_{\rm ion}$, and that furthermore $E_{\rm el}\gg E_{\rm
ion}$ ($\theta$ is small). Because $\xi\approx 2$, we know that half of
the gravitational potential energy is radiated away and half of it
goes into internal energy.  The problem is to decide how this energy
is split into an electronic and an ionic part.  The gravitational
energy changes with some average value of the interior density as
$E_{\rm g}\propto 1/R \propto \rho^{1/3}$. The energy of the
degenerate electrons is essentially the Fermi energy: $E_{\rm
el}\propto \rho^{2/3}$. Therefore, $\dot{E}_{\rm el}\approx 2(E_{\rm
el}/ E_{\rm g})\dot{E}_{\rm g}$. Using the virial theorem, this yields:
\begin{eqnarray}
\dot{E}_{\rm el}&\approx& -\dot{E}_{\rm g}\approx 2L \\
L &\approx& -\dot{E}_{\rm ion} \propto -\dot{T}.  
\end{eqnarray} 
The gravitational energy lost is entirely absorbed by the degenerate
electrons, and the observed luminosity is due to the thermal cooling
of the ions. 

Several simplifications limit the applicability of this result (that
would be valid in the white dwarf regime). In particular, the
coulomb and exchange terms in the EOS introduce negative
contributions that cannot be neglected. However, the approach is
useful to grasp how the evolution proceeds: in its
very early stages, the planet is very compressible. It follows a
standard Kelvin-Helmoltz contraction. When degeneracy sets in, the
compressibility becomes much smaller ($\alpha T\sim 0.1$, where
$\alpha$ is the coefficient of thermal expansion), and the planet
gets its luminosity mostly from the thermal cooling of the ions. The
luminosity can be written in terms of a modified Kelvin-Helmoltz
formula: 
\begin{equation}
L\approx \eta {GM^2\over R\tau},
\label{eq:lapprox}
\end{equation}
where $\tau$ is the age, and $\eta$ is a factor that hides most of the
complex physics. In the approximation that coulomb and exchange
terms can be neglected, $\eta\approx\theta/(\theta +1)$. The poor
compressibility of giant planets in their mature evolution stages
imply that $\eta\ll 1$ ($\eta\sim 0.03$ for Jupiter): the luminosity
is not obtained from the entire gravitational potential, but from the
much more limited reservoir constituted by the thermal internal
energy. Equation~\ref{eq:lapprox} shows that to first order, $\log
L\propto -\log\tau$: very little time is spent at high luminosity
values. In other words, the problem is (in most cases) weakly
sensitive to initial conditions. However, it is to be noticed
that with progresses in our capabilities to detect very young objects,
i.e. planets and brown dwarfs of only a few million years of age, the
problem of the initial conditions does become important \citep{Marley+06}.

\begin{figure}[p]
  \centerline{\resizebox{12cm}{!}{\includegraphics[angle=-90]{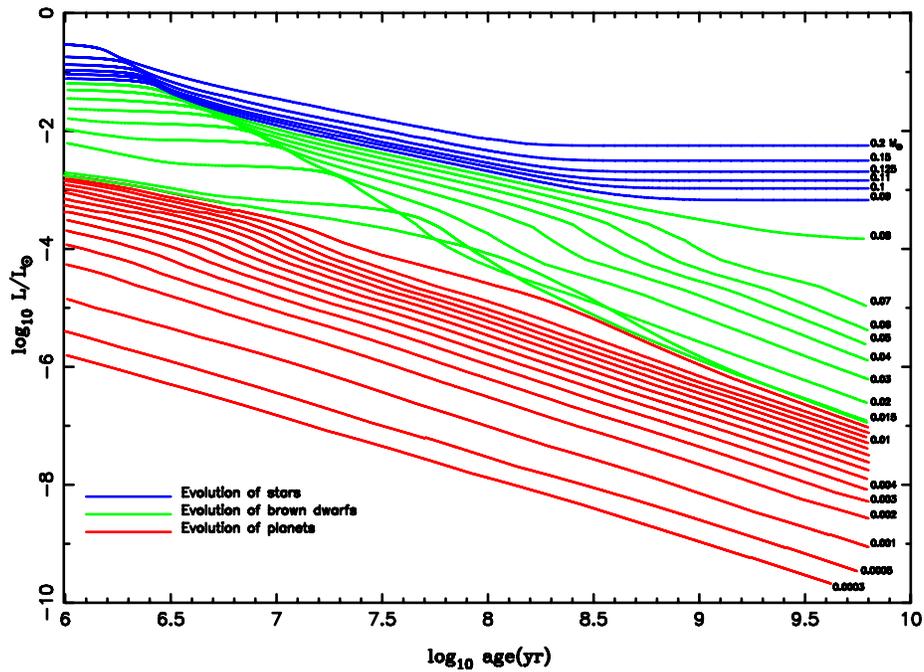}}}
\caption{Evolution of the luminosity (in L$_\odot$) of
solar-metallicity M dwarfs and substellar objects vs. time (in yr)
after formation. In this figure, "brown dwarfs" are arbitrarily
designated as those objects that burn deuterium, while those that do
not are tentatively labelled "planets". Stars are objects massive
enough to halt their contraction due to hydrogen fusion.  Each curve
is labelled by its corresponding mass in M$_\odot$, with the lowest three
corresponding to the mass of Saturn, half the mass of Jupiter, and the
mass of Jupiter.  [From \citet{Burrows+97}].}
\label{fig:L vs t}
\end{figure}

Figure~\ref{fig:L vs t} shows more generally how giant planets, but
also brown dwarfs and small stars see their luminosities evolve as a
function of time. The $1/\tau$ slope is globally conserved, with some
variations for brown dwarfs during the transient epoch of deuterium
burning, and of course for stars, when they begin burning efficiently
their hydrogen and settle on the main sequence: in that case, the
tendency of the star to contract under the action of gravity is
exactly balanced by thermonuclear hydrogen fusion.

\subsection{Mass-radius relation}

\begin{figure}[htbp]
\resizebox{12cm}{!}{\includegraphics[angle=0]{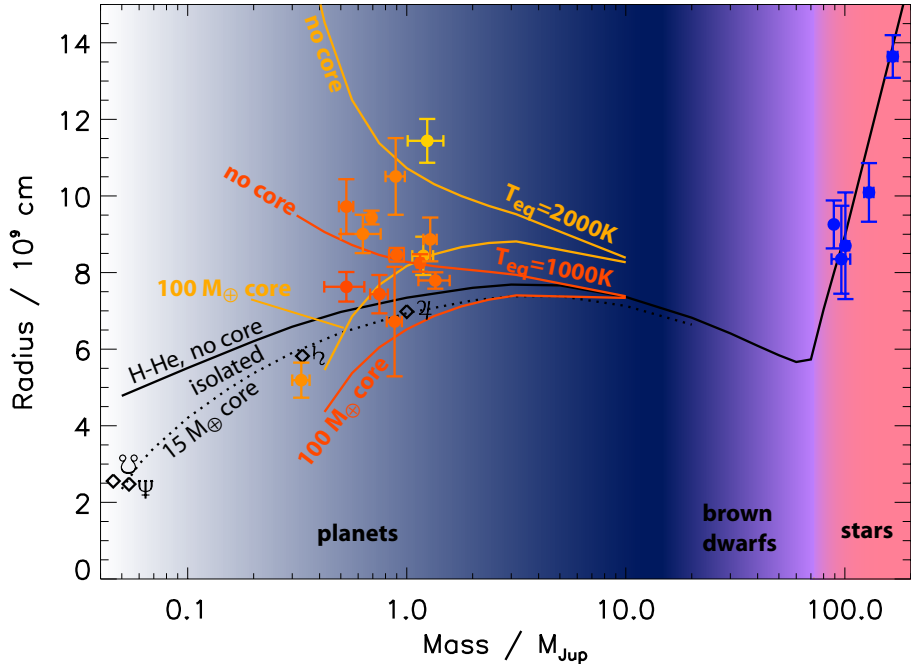}}
\caption{Theoretical and observed mass-radius relations. The black
  line is applicable to the evolution of solar composition planets,
  brown dwarfs and stars, when isolated or nearly isolated (as
  Jupiter, Saturn, Uranus and Neptune, defined by diamonds and their
  respective symbols), after 5 Ga of evolution. The dotted line shows
  the effect of a $15\mea$ core on the mass-radius relation. Orange
  and yellow curves represent the mass-radius relations for heavily
  irradiated planets with equilibrium temperatures of 1000 and
  2000\,K, respectively, and assuming that 0.5\% of the incoming
  stellar luminosity is dissipated at the center (see
  section~\ref{sec:irradiated}). For each irradiation level, two cases
  are considered: a solar-composition planet with no core (top curve),
  and one with a $100\mea$ central core (bottom curve). The transiting
  extrasolar giant planets for which a mass and a radius was measured
  are shown with points that are color-coded in function of the
  planet's equilibrium temperature. The masses and radii of very low
  mass stars are also indicated as blue points with error bars. 
}
\label{fig:mass_rad}
\end{figure}

The relation between mass and radius has very fundamental astrophysical
applications. Most importantly it allows one to infer the gross
composition of an object from a measurement of its mass and
radius. This is especially relevant in the context of the discovery 
of extrasolar planets with both radial velocimetry and the transit
method, as the two techniques yield relatively accurate determination
of $M$ and $R$. 

Figure~\ref{fig:mass_rad} shows mass-radius relations for compact
degenerate objects from giant planets to brown dwarfs and low-mass
stars. The right-hand side of the diagram shows a rapid increase of
the radius with mass in the stellar regime which is directly due to
the onset of stable thermonuclear reactions. In this regime,
observations and theoretical models agree \citep[see however][for a
more detailed discussion]{Ribas06}. The left-hand side of the diagram
is obviously more complex, and this can be understood by the fact that
planets have much larger variations in compositions than stars, and
because external factors such as the amount of irradiation they
receive do affect their contraction in a significant manner.

Let us first concentrate on isolated or nearly-isolated gaseous
planets. The black curves have a local maximum near $4\mjup$: at
small masses, the compression is small so that the radius
increases with mass. At large masses, degeneracy sets in and the
radius decreases with mass.

This can be understood on the basis of polytropic models based on the
assumption that $P=K\rho^{1+1/n}$, where $K$ and $n$ are
constants. Because of degeneracy, a planet of large mass will tend to
have $n\rightarrow 1.5$, while a planet a smaller mass will be less
compressible ($n\rightarrow 0$). Indeed, it can be shown that in their
inner 70 to 80\% in radius isolated solar composition planets of 10, 1
and $0.1\mjup$ have $n=1.3$, 1.0 and 0.6, respectively. From
polytropic equations \citep[e.g.][]{Chandrasekhar39}:
\begin{equation}
R\propto K^{n\over 3-n} M^{1-n\over 3-n}.
\label{eq:m-r-k}
\end{equation}
Assuming that $K$ is independant of mass, one gets $R\propto
M^{0.16}$, $M^{0}$, and $M^{-0.18}$ for $M=10$,
1 and $0.1\mjup$, respectively, in relatively good agreement with
\rfig{mass_rad} (the small discrepancies are due to the fact that the
intrinsic luminosity and hence $K$ depend on the mass considered).

Figure~\ref{fig:mass_rad} shows already that the planets in our Solar
System are not made of pure hydrogen and helium and require an
additional fraction of heavy elements in their interior, either in the
form of a core, or distributed in the envelope (dotted line). 

For extrasolar planets, the situation is complicated by the fact that
the intense irradiation that they receive plays a major role in their
evolution. The present sample is already quite diverse, with
equilibrium temperature (defined as the effective temperature
corresponding to the stellar flux received by the planet) ranging from
1000 to 2500\,K. Their composition is also quite variable, with some
planets having large masses of heavy elements
\citep{Sato+05,Guillot+06}. The orange and yellow curves in
fig.~\ref{fig:mass_rad} show theoretical results for equilibrium
temperatures of 1000 and 2000\,K, respectively. Two extreme models
have been plotted: assuming a purely solar composition planet (top
curve), and assuming the presence of a $100\mea$ central core (bottom
curve). In each case, an additional energy source proportional to
0.5\% of the incoming luminosity was also assumed (see discussion in
\S~\ref{sec:irradiated} hereafter).

The increase in radius for decreasing planetary mass for irradiated,
solar-composition planets with little or no core can be understood
using the polytropic relation (eq.~\ref{eq:m-r-k}), but accounting for
variations of $K$ as defined by the atmospheric boundary
condition. Using the Eddington approximation, assuming $\kappa\propto
P$ and a perfect gas relation in the atmosphere, one can show that
$K\propto (M/R^2)^{-1/2n}$ and that therefore $R\propto M^{1/2-n\over
2-n}$. With $n=1$, one finds $R\propto M^{-1/2}$. Strongly irradiated
hydrogen-helium planets of small masses are hence expected to have the
largest radii which qualitatively explain the positions of the
extrasolar planets in \rfig{mass_rad}.  Note that this estimate implicitly
assumes that $n$ is constant throughout the planet. The real situation
is more complex because of the growth of a deep radiative region in
most irradiated planets, and because of structural changes between the
degenerate interior and the perfect gas atmosphere.

In the case of the presence of a fixed mass of heavy elements, the
trend is inverse because of the increase of mean molecular mass (or
equivalently core/envelope mass) with decreasing total mass. Thus,
small planets with a core are much more tightly bound and less
subject to evaporation than those that have no core.

\subsection{Rotation and the figures of planets}
\label{sec:rotation}

The mass and radius of a planet informs us on its global
composition. Because planets are also rotating, one is allowed to
obtain more information on their deep interior structure. 
The hydrostatic equation becomes more complex however:
\begin{equation}
{\bfnab P\over \rho}=\bfnab\left(G\int\!\!\!\int\!\!\!\int 
{\rho(\bfr')\over |\bfr - \bfr'|}d^3\bfr'\right) - \bfOm\times(\bfOm\times\bfr),
\label{eq:full_hydrostat}
\end{equation}
where $\bfOm$ is the rotation vector. 
The resolution of eq.~(\ref{eq:full_hydrostat}) is a complex
problem. It can however be somewhat simplified by assuming that
$|\bfOm|\equiv\omega$ is such that the centrifugal force can be
derived from a potential. The hydrostatic equilibrium then writes
$\nabla P = \rho \nabla U$, and the {\it figure} of the rotating
planet is then defined by the $U=constant$ level surface. 

One can show \citep[e.g.][]{ZT78} that the hydrostatic
equation of a fluid planet can then be written in terms of the mean
radius $\rbar$ (the radius of a sphere containing the same volume as
that enclosed by the considered equipotential surface):
\begin{equation}
{1\over \rho}\dpar{P}{\rbar}=-{Gm\over \rbar^2}+{2\over 3}\omega^2
\rbar + {GM\over \overline{R}^3} \rbar\varphi_\omega,
\end{equation}
where $M$ and $\overline{R}$ are the total mass and mean radius of the
planet, and $\varphi_\omega$ is a slowly varying function of
$\rbar$. (In the case of Jupiter, $\varphi_\omega$ varies from about
$2\times 10^{-3}$ at the center to $4\times 10^{-3}$ at the surface.)
Equations~(\ref{eq:dtdr}-\ref{eq:dldr}) remain the same with the
hypothesis that the level surfaces for the pressure, temperature, and
luminosity are equipotentials.  The significance of rotation is
measured by the ratio of the centrifugal acceleration to the gravity:
\begin{equation}
q={\omega^2 \req^3\over GM}.
%;\qquad \qbar={\omega^2 \overline{R}^3\over GM}.
\end{equation}

%% The external gravitational potential of the planet is (assuming
%% hydrostatic equilibrium):
%% \begin{equation}
%% V_{\rm ext}(r,\cos\theta)={GM\over r}\left[1-\sum_{n=1}^\infty\left(a\over
%% r\right)^{2n}J_{2n}P_{2n}(\cos\theta)\right],
%% \end{equation}
%% where the coefficients $J_{2n}$ are the planet's {\it gravitational
%% moments}, and the $P_{2n}$ are Legendre polynomials. 

As discussed in section~\ref{sec:gravity}, in some cases, the external
gravity field of a planet can be accurately measured in the form of
gravitational moments $J_{k}$ (with zero odd moments for a planet in
hydrostatic equilibrium) that measure the departure from spherical
symmetry. Together with the mass, this provides a constraint on the
interior density profile (see \cite{ZT74} -see also
chapters by Van Hoolst and Sohl \& Schubert):
\begin{eqnarray*}
M&=&\int\!\!\!\int\!\!\!\int \rho(r,\theta) d^3\tau, \\
J_{2i} &=& -{1\over M R_{\rm eq}^{2i}}\int\!\!\!\int\!\!\!\int \rho(r,\theta) r^{2i}
P_{2i}(\cos\theta) d^3\tau,
\end{eqnarray*}
where $d\tau$ is a volume element and the integrals are performed over
the entire volume of the planet.

\begin{figure}[htbp]
\resizebox{10cm}{!}{\includegraphics[angle=0]{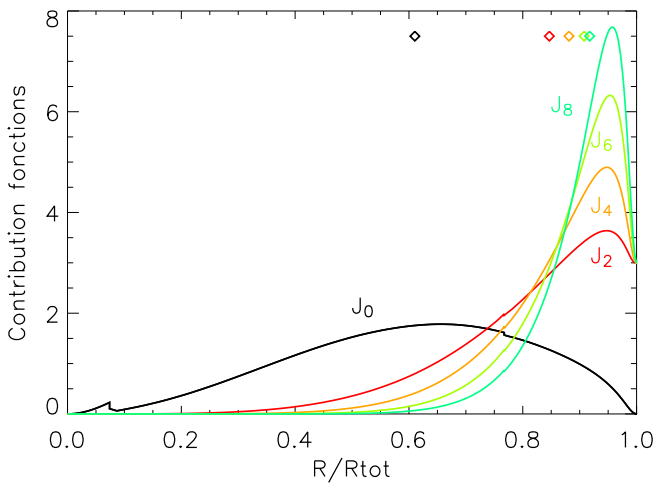}}
%\centerline{\hbox{\psfig{file=poly_index.ps,width=10cm}}}
\caption{Contribution of the level radii to the gravitational moments
  of Jupiter. $J_0$ is equivalent to the planet's mass. The small
  discontinuities are caused by the following transitions, from left
  to right: core/envelope, helium rich/helium poor
  (metallic/molecular). Diamonds indicate the median radius for each
  moment.}
\label{fig:contrib}
\end{figure}

Figure~\ref{fig:contrib} shows how the different layers inside a
planet contribute to the mass and the gravitational moments. The
figure applies to Jupiter, but would remain relatively similar for
other planets. Note however that in the case of Uranus and Neptune,
the core is a sizable fraction of the total planet and contributes
both to $J_2$ and $J_4$. Measured gravitational moments thus provide
information on the external levels of a planet. It is only indirectly,
through the constraints on the outer envelope that the presence of a
central core can be infered. As a consequence, it is impossible to
determine this core's state (liquid or solid), structure
(differentiated, partially mixed with the envelope) and composition
(rock, ice, helium...).

For planets outside the solar system, although measuring their
gravitational potential is utopic, their oblateness may be reachable
with future space transit observations \citep{SH02}. Since the
oblateness $e$ is, to first order, proportionnal to $q$:
\begin{equation}
e={\req\over\req-\rpol}\approx \left({3\over 2}\Lambda_2+{1\over 2}\right)q
\end{equation}
(where $\Lambda_2=J_2/q\approx 0.1$ to 0.2), it may be possible to
obtain their rotation rate, or with a rotation measured from another
method, a first constraint on their interior structure.

\section{Interior structures and evolutions}
\subsection{Jupiter and Saturn}\label{sec:JupSat}
\begin{figure}[htbp]
  \centerline{\resizebox{12cm}{!}{\includegraphics[angle=90,bb=4.5cm
	7cm 15cm 24cm]{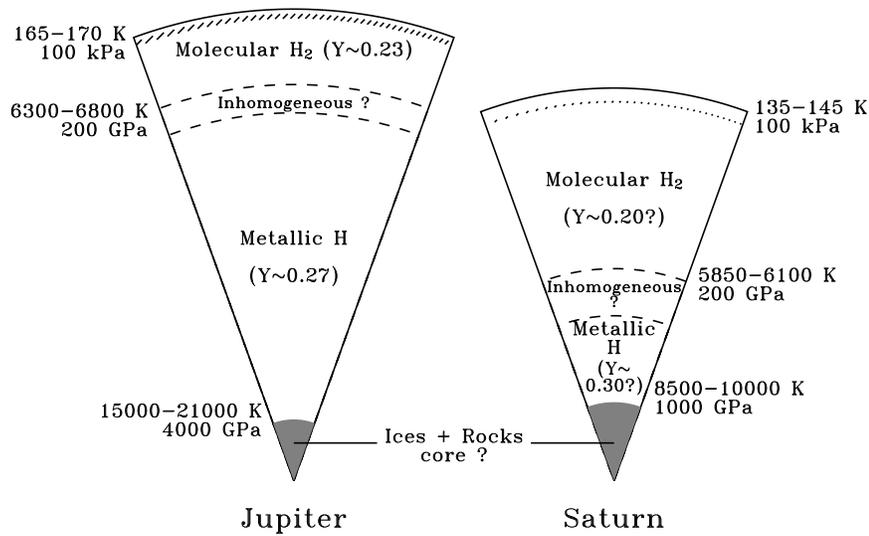}}}
\caption{Schematic representation of the interiors of Jupiter and
  Saturn. The range of temperatures is estimated using homogeneous
  models and including a possible radiative zone indicated by the hashed
  regions. Helium mass mixing ratios $Y$ are indicated. The size of the
  central rock and ice cores of Jupiter and Saturn is very uncertain
  (see text). In the case of Saturn, the inhomogeneous region may
  extend down all the way to the core which would imply the formation
  of a helium core. [Adapted from \citet{Guillot99b}].}
\label{fig:intjupsat}
\end{figure}

As illustrated by fig.~\ref{fig:intjupsat}, the simplest interior
models of Jupiter and Saturn matching all observational constraints
assume the presence of three main layers: (i) an outer hydrogen-helium
envelope, whose global composition is that of the deep atmosphere;
(ii) an inner hydrogen-helium envelope, enriched in helium because the
whole planet has to fit the H/He protosolar value; (iii) a central
dense core. Because the planets are believed to be mostly convective,
these regions are expected to be globally homogeneous. (Many
interesting thermochemical transformations take place in the deep
atmosphere, but they are of little concern to
us). 

The transition from a helium-poor upper envelope to a helium-rich
lower envelope is thought to take place through the formation of
helium-rich droplets that fall deeper into the planet due to their
larger density. These droplets form because of an assumed phase
transition of helium in hydrogen at high pressures and low
temperatures. Three-layer models implicitely make the hypothesis that
this region is narrow. Indeed, calculations of such a phase separation
in a fully-ionized plasma indicate a rapid decrease of the critical
temperature with increasing pressure, with the consequence that helium
would be unsoluble in a relatively small region in low-pressure
metallic hydrogen. This region would progressively grow with time
\citep[e.g.][]{Stevenson82}. However, DFT calculations have indicated that
the critical temperature for helium demixing may rise with pressure
\citep{PHB95}, presumably in the regime where hydrogen
is only partially ionized and bound states remain. This opens up the
possibility that the inhomogeneous regions may be more extended, and
that models more complex than the three-layer models may be needed, in
particular in the case of Saturn (see below).

In the absence of these calculations, the three-layer models can be
used as a useful guidance to a necessarily hypothetical ensemble of
allowed structures and compositions of Jupiter and Saturn. These
relatively extensive exploration of the parameter space have been
performed by \citet{SG04}. The calculations assume that
only helium is inhomogeneous in the envelope (the abundance of heavy
elements is supposed to be uniform accross the molecular/metallic
hydrogen transition). Many sources of uncertainties are taken into
account however; among them, the most significant are on the equations
of state of hydrogen and helium, the uncertain values of $J_4$ and
$J_6$, the presence of differential rotation deep inside the planet,
the location of the helium-poor to helium-rich region, and the
uncertain helium to hydrogen protosolar ratio.

Their results indicate that Jupiter's core is smaller than $\sim
10\mea$, and that its global composition is pretty much unknown
(between 10 to 42$\mea$ of heavy elements in total). The models
indicate that Jupiter is enriched compared to the solar value by a
factor 1.5 to 8 times the solar value. This enrichment is compatible
with a global uniform enrichment of all species near the atmospheric
Galileo values, but include many other possibilities.

%% Most of the constraints are derived from the values of the radius (or
%% equivalently mass) and of $J_2$. The measurement of $J_4$ allows to
%% further narrow the ensemble of possible models, and in some cases, to
%% rule out EOS solutions (in particular those indicating relatively
%% large core masses, between 10 and 20$\mea$).  As discussed in Guillot
%% (1999a) and Saumon \& Guillot (2004), most of the uncertainty in the
%% solution arises because very different hydrogen EOSs are possible. The
%% fact that more laboratory and numerical experiments seem to indicate
%% relatively low-compressions for hydrogen at Mbar pressures points
%% towards smaller core masses and a larger amount of heavy elements in
%% the planet. However, this relies on uncertain temperature gradients,
%% because the EOSs are based on laboratory data obtained at temperatures
%% higher than those relevant to the planetary interiors.

In the case of Saturn, the solutions depend less on the hydrogen EOS
because the Mbar pressure region is comparatively smaller. The total
amount of heavy elements present in the planet can therefore be
estimated with a better accuracy than for Jupiter, and is between $20$
and $30\mea$. In three-layer models with a discontinuity of the helium
abundance at the molecular-metallic hydrogen interface but continuity
of all other elements, the core masses found are between $10$ and
$22\mea$. However, because Saturn's metallic region is deeper into
the planet, it mimics the effect that a central core would have on
$J_2$. If we allow for variations in the abundance of heavy elements
together with the helium discontinuity, then the core mass can become
much smaller, and even solutions with no core can be found (Guillot
1999a). These solutions depend on the hypothetic phase separation of
an abundant species (e.g. water), and generally cause an energy
problem because of the release of considerable gravitational energy.
However, another possibility is through the formation of an almost
pure helium shell around the central core, which could lower the core
masses by up to $~7\mea$ \citep[][Hubbard, personnal
communication]{FH03}.

\begin{figure}[htb]
\centerline{\resizebox{8cm}{!}{\includegraphics[angle=0]{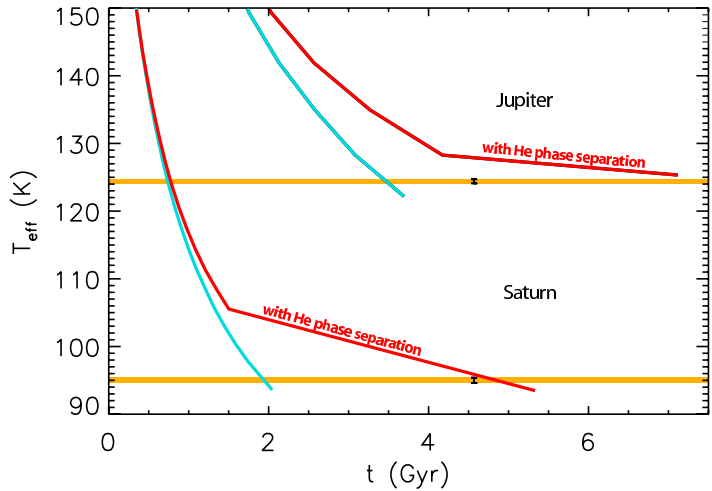}}}
\vspace*{1cm}
\caption{Final stages of evolution of Jupiter and Saturn. The present
  effective temperatures, reached after $\sim 4.55$\,Ga of
  evolution, are indicated as horizontal orange lines. For each planet
  two models represent attempts to bracket the ensemble of
  possibilities, with the faster evolution corresponding to that of an
  homogeneous planet, while the slowest evolution includes the effect
  of helium settling in the last evolution phase. [Adapted from
  \cite{Hubbard+99} and \cite{FH03}].}
\label{fig:jup-sat-evolution}
\end{figure}

Concerning the {\it evolutions} of Jupiter and Saturn, the three main
sources of uncertainty are, by order of importance: (1) the magnitude
of the helium separation; (2) the EOS; (3) the atmospheric boundary
conditions. Figure~\ref{fig:jup-sat-evolution} shows an ensemble of
possibilities that attempt to bracket the minimum and maximum
cooling. In all cases, helium sedimentation is needed to explain
Saturn's present luminosity \citep[see][]{Salpeter73, SS77b,
Hubbard77}. Recent models of Saturn's evolution appear to favor a
scenario in which helium settles down almost to the central core
\citep{Hubbard+99,FH03}. In the case of Jupiter, the sedimentation of
helium that appears to be necessary to explain the low atmospheric
helium abundance poses a problem for evolution models because it
appears to generally prolong its evolution beyond 4.55\,Ga, the age of
the Solar System. However, different solutions are possible, including
improvements of the EOS and atmospheric boundary conditions, or even
the possible progressive erosion of the central core that would yield
a lower Jupiter's luminosity at a given age \citep{GSHS04}.

\subsection{Uranus and Neptune}
\begin{figure}[htbp]
\centerline{\resizebox{11cm}{!}{\includegraphics[angle=90,bb=4.5cm 4.5cm 15cm 24cm]{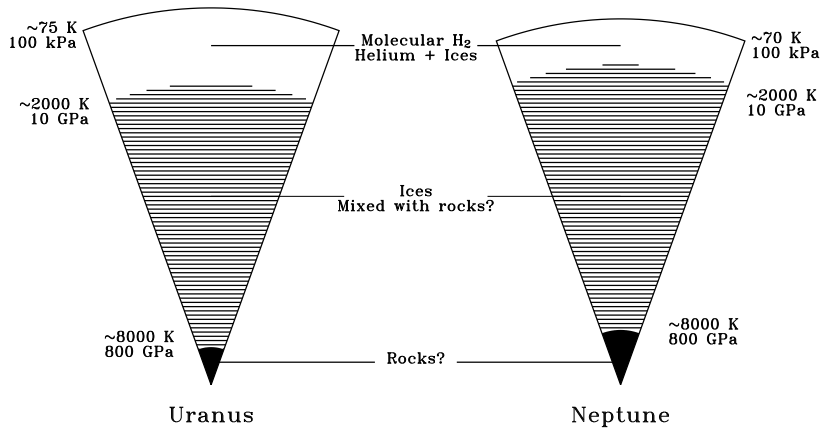}}}
%\vspace*{-1cm}
\caption{Schematic representation of the interiors of Uranus and
  Neptune. [Adapted from \cite{Guillot99b}].}
\label{fig:inturanep}
\end{figure}

Although the two planets are relatively similar, \rfig{mass_rad}\ already
shows that Neptune's larger mean density compared to Uranus has to be
due to a slightly different composition: either more heavy elements
compared to hydrogen and helium, or a larger rock/ice ratio.  The
gravitational moments impose that the density profiles lie close to
that of ``ices'' (a mixture initially composed of e.g. H$_2$O, CH$_4$ and
NH$_3$, but which rapidly becomes a ionic fluid of uncertain chemical
composition in the planetary interior), except in the outermost
layers, which have a density closer to that of hydrogen and helium
\citep{MGP95,PPM00}. As illustrated in
\rfig{inturanep}, three-layer models of Uranus and Neptune consisting
of a central ``rocks'' core (magnesium-silicate and iron material), an
ice layer and a hydrogen-helium gas envelope have been calculated
\citep{PHS91, Hubbard+95}.

The fact that models of Uranus assuming homogeneity of each layer and
adiabatic temperature profiles fail in reproducing its gravitational
moments seem to imply that substantial parts of the planetary interior
are not homogeneously mixed \citep{PWM95}.  This could explain
the fact that Uranus' heat flux is so small: its heat would not be
allowed to escape to space by convection, but through a much slower
diffusive process in the regions of high molecular weight
gradient. Such regions would also be present in Neptune, but much
deeper, thus allowing more heat to be transported outward.  The
existence of these non-homogeneous, partially mixed regions are
further confirmed by the fact that if hydrogen is supposed to be
confined solely to the hydrogen-helium envelope, models predict
ice/rock ratios of the order of 10 or more, much larger than the
protosolar value of $\sim\,$2.5. On the other hand, if we impose the
constraint that the ice/rock ratio is protosolar, the overall
composition of both Uranus and Neptune is, by mass, about 25\% rock,
$60-70$\% ice, and $5-15$\% hydrogen and helium \citep{PHS91,
PWM95, Hubbard+95}. Assuming both ice and rock are present
in the envelope, an upper limit to the amount of hydrogen and helium
present is $\sim 4.2\mea$ for Uranus and $\sim 3.2\mea$ for Neptune
\citep{PPM00}. A lower limit of $\sim 0.5\mea$ for both
planets can be inferred by assuming that hydrogen and helium are only
present in the outer envelope at $P\wig<100$\,kbar.

\subsection{Irradiated giant planets}
\label{sec:irradiated}

Although all extrasolar giant planets are in principle interesting, we
focus here on the ones that orbit extremely close to their star
because of the possibility to directly characterise them and measure
their mass, radius and some properties of their atmosphere. Two
planets are proxies for this new class of objects: the first
extrasolar giant planet discovered, 51\,Peg\,b, with an orbital period
of $P=4.23$ days, and the first {\em transiting\/} extrasolar giant
planet, HD\,209458\,b, with $P=3.52$ days. Both planets belong to the
Pegasus constellation, and following astronomical conventions
(e.g. Cepheids, named after $\delta$~Cephei), we choose to name giant
planets orbiting close to their stars with periods shorter than 10
days ``Pegasids'' (alternatively, ``hot Jupiters'' is also found in
the litterature).

With such a short orbital period, these planets are for most of them
subject to an irradiation from their central star that is so intense
that the absorbed stellar energy flux can be about $\sim 10^4$ times
larger than their intrinsic flux. The atmosphere is thus prevented
from cooling, with the consequence that a radiative zone develops and
governs the cooling and contraction of the interior
\citep{Guillot+96}. Typically, for a planet like HD\,209458\,b, this
radiative 
zone extends to kbar levels, $T\sim 4000\K$, and is located in the
outer 5\% in radius ($0.3\%$ in mass) \citep{GS02}.

\begin{figure}[htbp]
\begin{center}
  \centerline{\resizebox{10cm}{!}{\includegraphics[angle=0]{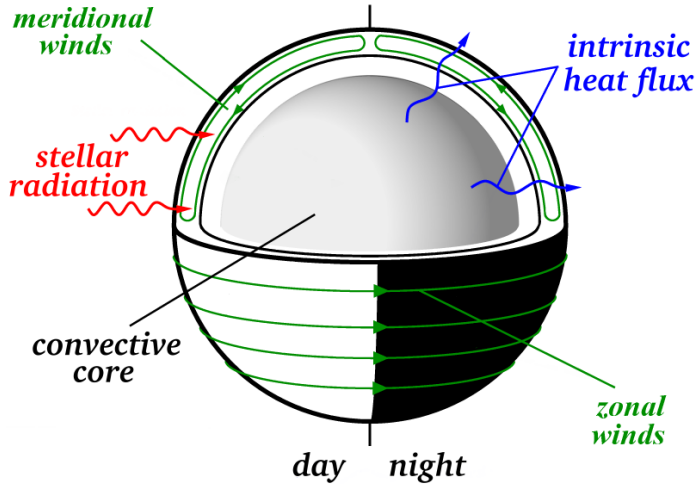}}}
\caption{Conjectured dynamical structure of Pegasids (strongly
  irradiated extrasolar giant planets): At pressures larger than
  100--$800$\,bar, the intrinsic heat flux must be transported by
  convection. The convective core is at or near synchronous rotation
  with the star and has small latitudinal and longitudinal temperature
  variations. At lower pressures a radiative envelope is present. The
  top part of the atmosphere is penetrated by the stellar light on the
  day side. The spatial variation in insolation should drive winds
  that transport heat from the day side to the night side. [From
  \cite{SG02}].}
\label{fig:circ}
\end{center}
\end{figure}

\begin{figure}[htbp]
\begin{center}
\centerline{\resizebox{12cm}{!}{\includegraphics[angle=0]{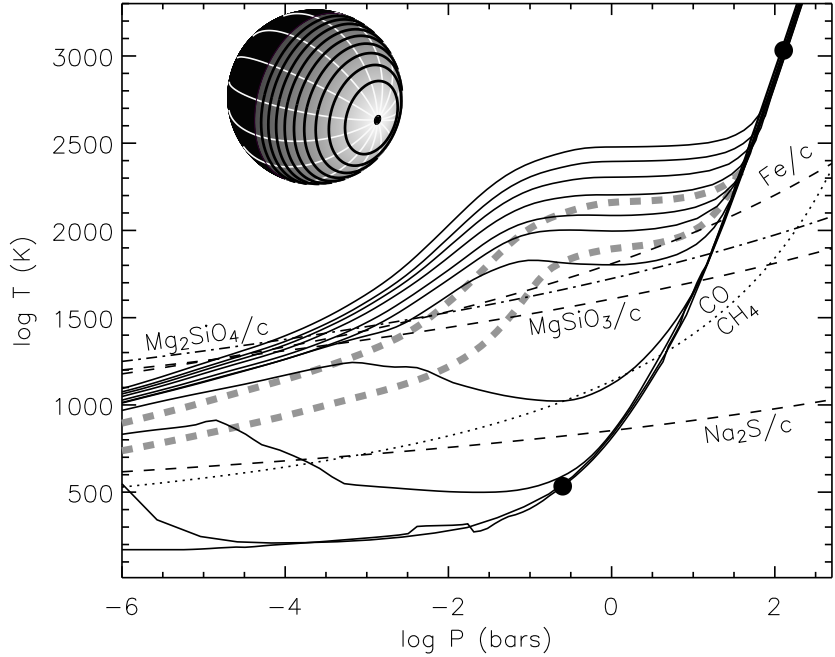}}}
\caption{Temperature versus pressure for a sequence of locations in
  the atmosphere of HD209458b, assuming no horizontal redistribution
  of heat. Each sequence corresponds to a given direction of the
  incident flux relative to the surface normal. The approximate
  regions represented by the collection of T-P profiles are shown as
  solid black lines on the illustrative sphere.  The top-most T-P
  profile corresponds to the sub-stellar point (black dot on the
  sphere).  The terminator and night side (black hemisphere) are
  modeled with the non-irradiated profile (lowest T-P curve).  The
  radiative-convective boundary at the sub-stellar point and on the
  night side are labeled with filled circles. The dashed lines
  indicate the approximate condensation curves for three common grain
  species.  The dotted line indicates where gaseous CO and CH$_4$
  concentrations are equal (CO is dominant to the left of this line).
  The thick, grey, dashed lines are T-P profiles calculated for a
  normal incident flux equal to 0.5 (top) and 0.25 (bottom) times that
  at the substellar point, as often used as approximate solutions for
  the day side, or entire atmosphere, respectively. [From \cite{BHA05}].}
\label{fig:barman}
\end{center}
\end{figure}

Problems in the modeling of the evolution of Pegasids arise mostly
because of the uncertain outer boundary condition. The intense stellar
flux implies that the atmospheric temperature profile is extremely
dependant upon the opacity sources considered. Depending on the chosen
composition, the opacity data used, the assumed presence of clouds,
the geometry considered, resulting temperatures in the deep atmosphere
can differ by up to $\sim 600\K$ \citep{SS00, Goukenleuque+00, BHA01,
SBH03, IBG05, Fortney+06}. Furthermore, as illustrated by \rfig{circ},
the strong irradiation and expected synchronisation of the planets
implies that strong inhomogeneities should exist in the atmosphere
with in particular strong ($\sim 500$\,K) day-night and
equator-to-pole differences in effective temperatures \citep{SG02,
IBG05, CS05, BHA05}, further complicating the modeling of the
planetary evolution (see fig.~\rfig{barman}). Finally, another related
problem is the presence of the radiative zone. Again, the composition
is unknown and the opacity data are uncertain in this relatively high
temperature ($T\sim 1500-3000\K$) and high pressure (up to $\sim
1\,$kbar) regime.

%% \begin{figure}[htb]
%% \centerline{\resizebox{12cm}{!}{\includegraphics[angle=0]{ev-cold-col.ps}}}
%% \caption{Evolution of HD209458b using a ``cold'' atmospheric boundary
%% condition (see text).  The evolution of the central
%% pressure with time
%% is shown as the bottom thick line. The planet is convective except for
%% an upper radiative zone indicated by a hashed area. Isotherms from
%% 4000 to 20\,000\,K are indicated. The isotherms not labeled correspond
%% to 3500, 30\,000 and 40\,000\,K. The dashed line indicates the time
%% necessary to contract the planet to a radius of 1.35\,\rjup.
%% [From Guillot \& Showman (2002)]}
%% \label{fig:ev-cold}
%% \end{figure}

We have seen in fig.~\rfig{mass_rad} that the measured masses and
radii of transiting planets can be globally explained in the framework
of an evolution model including the strong stellar irradiation and the
presence of a variable mass of heavy elements, either in the form of a
central core, or spread in the planet interior. However, when
analyzing the situation for each planet, it appears that several
planets are too large to be reproduced by standard models, i.e. models
using the most up-to-date equations of state, opacities, atmospheric
boundary conditions and assuming that the planetary luminosity
governing its cooling is taken solely from the lost gravitational
potential energy (see Section~\ref{sec:basic}).

\begin{figure}[htbp]
\centerline{\resizebox{12cm}{!}{\includegraphics[angle=0]{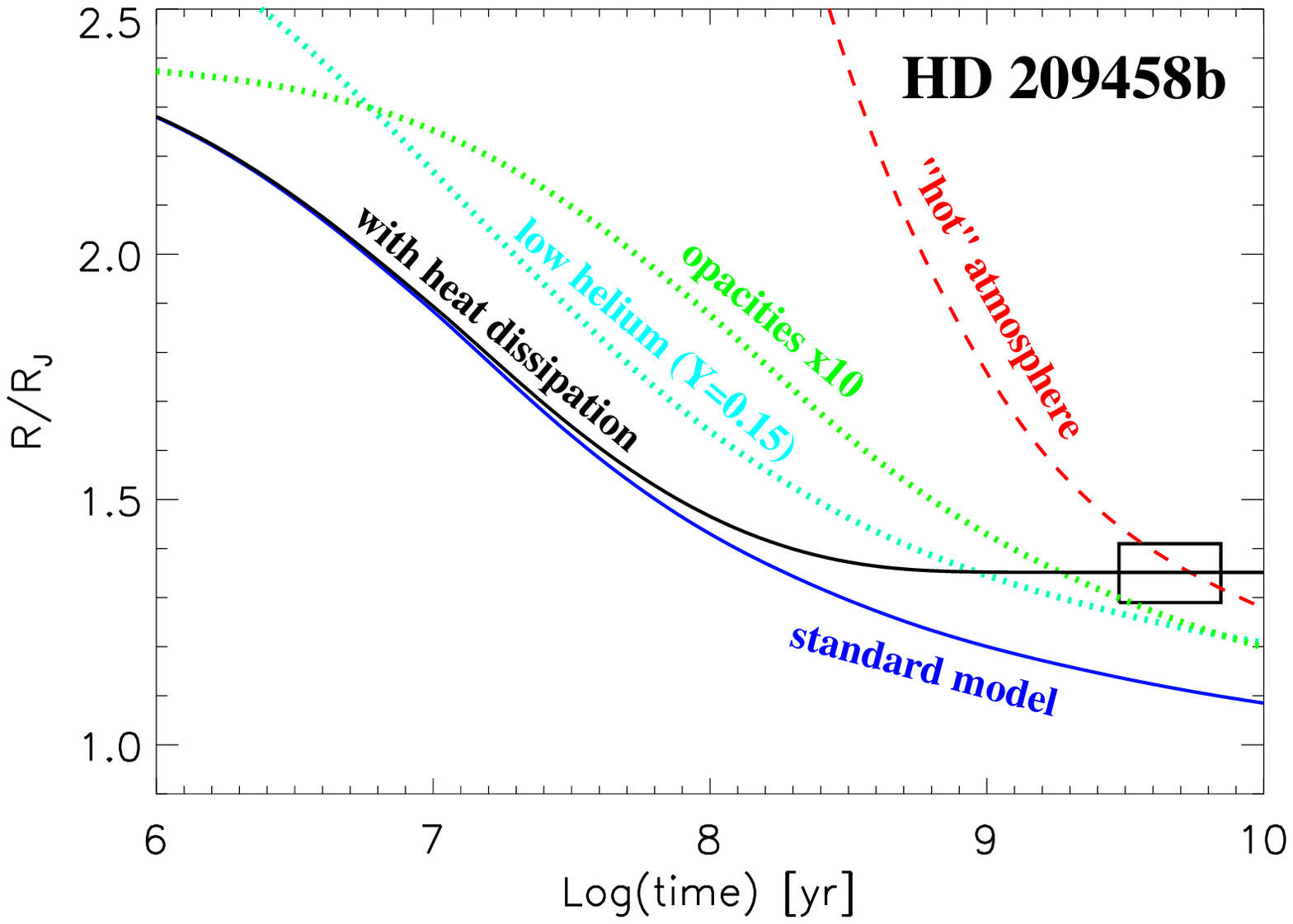}}}
\caption{The contraction of HD209458b as a function of time can
be compared to its measured radius and inferred age shown by the black
box. Standard models (blue curve) for the evolution of that $0.69\,\rm
M_{\rm J}$ planet generally yield a radius that is too small compared
to the observations, even for a solar composition and no central core
(a larger core and -in most cases- larger amounts of heavy elements in
the planet imply an even smaller size for a given
age). Unrealistically low helium abundances or high opacities models
lead to evolution tracks that barely cross the observational box. A
possiblity is that heat is dissipated into the deep interior by
stellar tides, either related to a non-zero orbital eccentricity
forced by an unseen companion, or because of a constant transfer of
angular momentum from the heated atmosphere to the interior (black
curve). Alternatively, the atmosphere may be hotter than predicted due
to heating by strong zonal winds and shear instabilities (red curve).}
\label{fig:ev-hd209458b}
\end{figure}

Figure~\ref{fig:ev-hd209458b} illustrates the situation for the
particular case of HD209458b: unless using an unrealistically hot
atmosphere, or arbitrarily increasing the internal opacity, or
decreasing the helium content, one cannot reproduce the observed
radius which is 10 to 20\% larger than calculated \citep{BLM01, BLL03,
  GS02, Baraffe+03}. The
fact that the measured radius corresponds to a low-pressure
($\sim$mbar) level while the calculated radius corresponds to a level
near 1\,bar is not negligible \citep{BSH03} but too small to
account for the difference. This is problematic because while it is
easy to invoke the presence of a massive core to explain the small
size of a planet, a large size such as that of HD209458b requires
an additional energy source, or significant modifications in the
data/physics involved.

\citet{BLM01} proposed that this large radius may be due
to a small forced eccentricity ($e\sim 0.03$) of HD209458b, and
subsequent tidal dissipation in the planet interior, but detailed
observations indicate that the eccentricity is small, $e=0.014\pm
0.009$ \citep{Laughlinetal2005b}, and observations of the secondary
eclipse imply that this would further require a chance configuration
of the orbit \citep{DSRH05}. Another proposed explanation also
involving tidal dissipation of orbital energy is that the planet may
be trapped in a Cassini state with a large orbital inclination
\citep{WinnHolman2005}, but it appears to have a low probability of
occurrence \citep{Levrard+06}. Finally, a third possibility that
would apply to {\it all} Pegasids is to invoke a downward transport of
kinetic energy and its dissipation by tides
\citep{SG02}. This last possibility would require the
various transiting planets to have different core masses to reproduce
the observed radii \citep{Guillot05}.

\begin{figure}
\centerline{\resizebox{10cm}{!}{\includegraphics{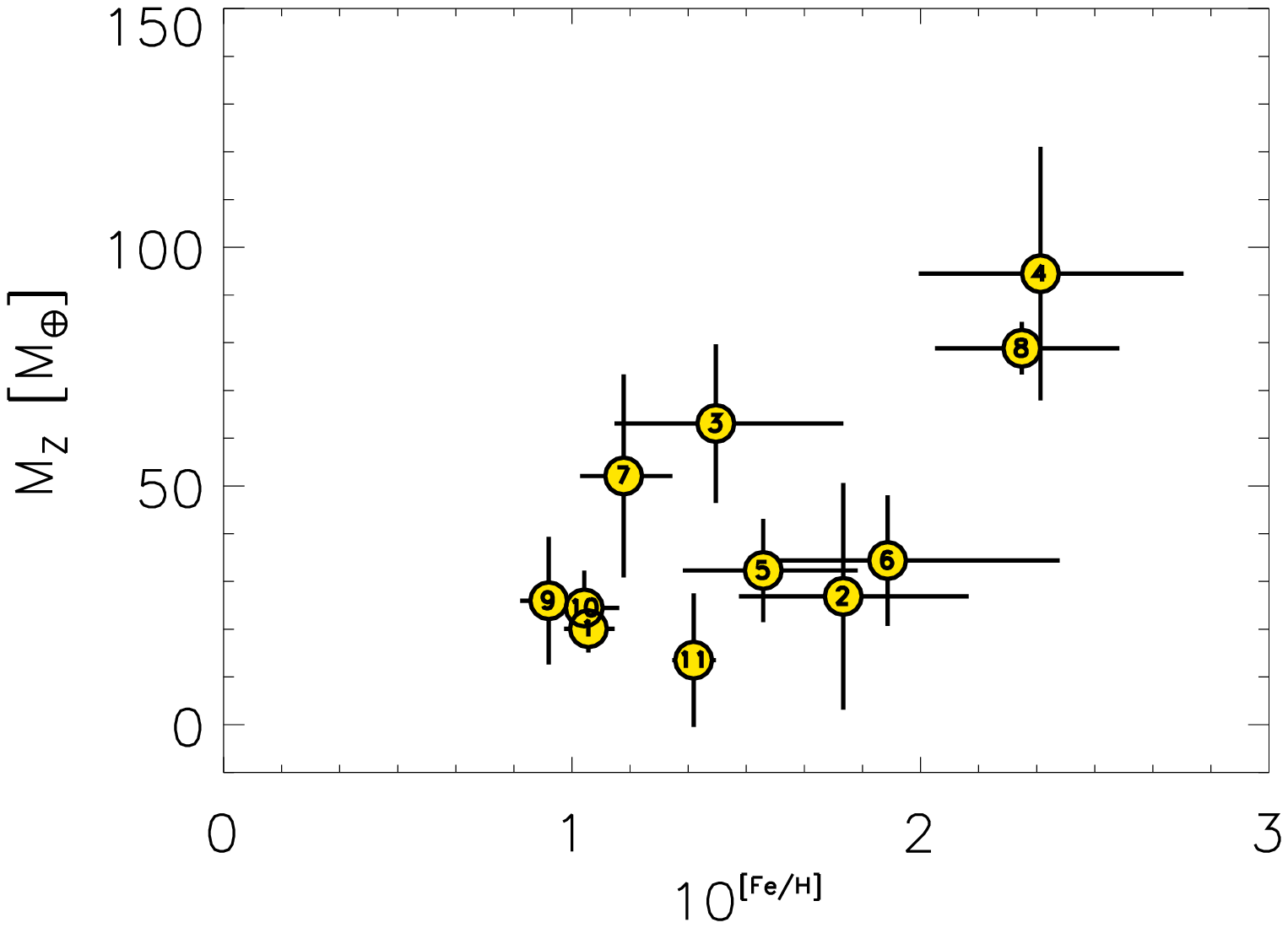}}}
\caption{Mass of heavy elements in transiting Pegasids known by 2006
  as a function of the metal content of the parent star relative to
  the Sun. The mass of heavy elements required to fit the measured
  radii is calculated on the basis of evolution models including an
  additional heat source slowing the cooling of the planet. This heat
  source is assumed equal to $0.5\%$ of the incoming stellar heat flux
  \citep{SG02}. Horizontal error bars correspond to the
  $1\sigma$ errors on the [Fe/H] determination. Vertical error bars
  are a consequence of the uncertainties on the measured planetary
  radii and ages. Note that the results, based on \citet{Guillot+06}
  are intrinsically model-dependent and may be affected by further
  discoveries of transiting planets.}
\label{fig:correlation}
\end{figure}

Recently, as more transiting Pegasids have been discovered, the number
of anomalously large ones has increased to at least 3 for 11 planets,
implying that this is not a rare event. This lends more weight to a
mechanism that would apply to each planet. In this case, masses of
heavy elements can be derived by imposing that all planets should be
fitted by the same model with the same hypotheses. This can be done by
inverting the results of \rfig{mass_rad}, as described by
\citep{Guillot+06}. The method is applied to the known transiting
Pegasids by the end of 2006 in \rfig{correlation}, a plot of the
masses of heavy elements in the planets as a function of the
metallicities of the parent star (which measures how rich a given star
is in heavy elements compared to the Sun).

Figure~\ref{fig:correlation} first shows that in some cases, large
masses of heavy elements (up to $\sim 100\mea$ are necessary. This is
in harmony with the composition inferred for HD149026b, i.e. around
$70\mea$ of heavy elements, a conclusion that is hard to escape
because of the low total mass and high irradiation of the planet
\citep[see][]{Ikoma+06, Fortney+06}.  Furthermore, there seems to be a
correlation between the mass of heavy elements inferred in the
transiting planets, and the metallicity of the parent stars
\citep{Guillot+06}, although this correlation has to be ascertained by
more measurements. A caveat is important: these results are
intrinsically model-dependent as they are based on the assumption that
all planets receive an additionnal tidal heat flux that is
proportionnal to the energy flux that they receive in the form of
photons from the star. More transiting planets are needed to confirm
or infirm this model, but the large variety in core masses, and the
absence of Pegasids around metal-poor stars ($\rm [Fe/H]\wig{<}
-0.07$) as indicated by \rfig{correlation} appear to be robust
consequences of this work.

Another intriguing possibility concerning Pegasids is that of a
sustained mass loss due to the high irradiation dose that the planets
receive. Indeed, this effect was predicted \citep{BL95, Guillot+96,
Lammer+03} and detected \citep{VidalMadjar+03, VidalMadjar+04}, but
its magnitude is still quite uncertain, by at least two orders of
magnitude \citep{Lammer+03, Lecavelier+04, Yelle06}. The effect on the
evolution is surprisingly limited, except at the final stages when an
exponential mass loss appears in fully gaseous planets
\citep{Baraffe+04}.

Finally, it is important to note that another class of planets awaits
a direct characterisation by the transit method: that of ice or rock
giants. Small-mass planets around $10\mea$ have been detected
\citep[e.g.][]{Lovis+06, Beaulieu+06} but their radius is
expected to be small \citep{Guillot+96, VOS06}, and we
currently may not have the observational capability to test whether
they transit in front of their star. This should be resolved by the
space mission CoRoT (launched on 27 dec 2006) and Kepler (launch $\sim
2008$). These objects are especially interesting but pose difficult
problems in terms of structure because depending on their formation
history, precise composition and location, they may be fluid, solid,
or they may even possess a global liquid ocean \citep[see][]{Kuchner03,
  Leger+04}.

\section{Implications for planetary formation models}

The giant planets in our Solar System have in common possessing a
large mass of hydrogen and helium, but they are obviously quite
different in their aspect and in their internal structures. Although
studies cannot be conducted with the same level of details, we can
safely conclude that extrasolar planets show an even greater variety
in composition and visible appearance. 

A parallel study of the structures of our giant planets and of giant
planets orbiting around other stars should provide us with key
information regarding planet formation in the next decade or so. But,
already, some conclusions, some of them robust, others still
tentative, can be drawn (see also the chapter by Stevenson):

{\it Giant planets formed in circumstellar disks, before these were
  completely dissipated:}\\
This is a relatively obvious consequence of the fact that giant
  planets are mostly made of hydrogen and helium: these elements had
  to be acquired when they were still present in the disk. Because the
  observed lifetime of gaseous circumstellar disks is of the order of
  a few million years, this implies that these planets formed
  (i.e. acquired most of their final masses) in a few million years
  also, quite faster than terrestrial planets in the Solar System. 

{\it Giant planets migrated:}\\
Although not cleanly demonstrated yet, there is evidence that the
observed orbital distribution of extrasolar planets requires an inward
migration of planets, and various mechanisms have been proposed for
that \citep[see][...etc.]{IL04a, AMBW05, MA05}. Separately, it was
shown that several properties of our 
Solar System can be explained if Jupiter, Saturn, Uranus and Neptune
ended up the early formation phase in the presence of a disk with
quasi-circular orbit, and with Saturn, Uranus and Neptune
significantly closer to the Sun than they are now, and that these
three planets subsequently migrated outward \citep{TGML05}.

{\it Accretion played a key role for giant planet formation:}\\
Several indications point towards a formation of giant planets that is
dominated by accretion of heavy elements: First, Jupiter, Saturn,
Uranus and Neptune are all significantly enriched in heavy elements
compared to the Sun. This feature can be reproduced by core-accretion
models, for Jupiter and Saturn at least \citep{AMBW05}. Second,
the probability to find a giant planet around a solar-type star (with
stellar type F, G or K) is a strongly rising function of stellar
metallicity \citep{Gonzalez98, SIM04, FV05}, a property
that is also well-reproduced by standard core accretion models
\citep{IL04b, AMBW05}. Third, the large masses of heavy
elements inferred in some transiting extrasolar planets as well as the
apparent correlation between mass of heavy elements in the planet and
stellar metallicity (\cite{Guillot+06}; see also \cite{Sato+05}
and \cite{Ikoma+06}) is a strong indication that accretion was
possible and that it was furthermore efficient. It is to be noted that
none of these key properties are directly explained by formation
models that assume a direct gravitational collapse \citep[see][]{Boss04, MQWS04}. 

{\it Giant planets were enriched in heavy elements by core accretion,
  planetesimal delivery and/or formation in an enriched protoplanetary
  disk:}\\
The giant planets in our Solar System are unambigously enriched in
  heavy elements compared to the Sun, both globally, and when
  considering their atmosphere. This may also be the case of
  extrasolar planets, although the evidence is still tenuous.
  The accretion of a central core can explain part of the global
  enrichment, but not that of the atmosphere. The accretion of
  planetesimals may be a possible solution but in the case of Jupiter
  at least the rapid drop in accretion efficiency as the planet
  reaches appreciable masses ($\sim 100\mea$ or so) implies that such
  an enrichment would have originally concerned only very deep layers,
  and would require a relatively efficient upper mixing of these
  elements, and possibly an erosion of the central core \citep{GSHS04}. 

Although not unambiguously explained, the fact that Jupiter is also
enriched in noble gases compared to the Sun is a key observation to
understand some of the processes occuring in the early Solar
System. Indeed, noble gases are trapped into solids only at very low
temperatures, and this tells us either that most of the solids that
formed Jupiter were formed at very low temperature to be able to trap
gases such as Argon, probably as clathrates \citep{GHML01, HGL04}, or
that the planet formed in an enriched disk as 
it was being evaporated \citep{GH06}.

\section{Future prospects}

We have shown that the compositions and structures of giant planets
remain very uncertain. This is an important problem when attempting
to understand and constrain the formation of planets, and the origins
of the Solar System. However, the parallel study of giant planets in
our Solar System by space missions such as Galileo and Cassini, and
of extrasolar planets by both ground based and space programs has led
to rapid improvements in the field, with in particular a precise
determination of the composition of Jupiter's troposphere, and
constraints on the compositions of a dozen of extrasolar planets.

Improvements on our knowledge of the giant planets requires a variety
of efforts. Fortunately, nearly all of these are addressed at least
partially by adequate projects in the next few years. The efforts that
are necessary thus include (but are not limited to):
\begin{itemize}
\item Obtain a better EOS of hydrogen, in particular near the
  molecular/metallic transition. This will be addessed by the
  construction of powerful lasers such as the NIF in the US and
  the M\'egaJoule laser in France, and by innovative experiments such
  as shocks on pre-compressed samples. One of the challenges is not only
  obtaining higher pressures, but mostly lower temperatures than
  currently possible with single shocks. The parallel improvement
  of computing facilities should allow more extended numerical
  experiments.
\item Calculate hydrogen-helium and hydrogen-water phase
  diagrams. (Other phase diagrams are desirable too, but of lesser
  immediate importance). This should be possible with new numerical
  experiments. 
\item Have a better yardstick to measure solar and protosolar
  compositions. This may be addressed by the analysis of the Genesis
  mission samples, or may require another future mission. 
\item Improve the values of $J_4$ and $J_6$ for Saturn. This will be
  done as part of the Cassini-Huygens mission. This should lead to
  better constraints, and possibly a determination of whether the
  interior of Saturn rotates as a solid body. 
\item Detect new transiting extrasolar planets, and hopefully some
  that are further from their star. The space missions CoRoT (2006)
  and Kepler (2008) should provide the detection and
  characterization of many tens, possibly hundreds of giant planets.
\item Model the formation and evolution of ice giants such as
  Uranus, Neptune, and similar planets around other stars, in order to
  analyze detections of these objects and understand planetary
  formation.
\item Improve the measurement of Jupiter's gravity field, and
  determine the abundance of water in the deep atmosphere. This will
  be done by the Juno mission (launch 2011) with a combination of an
  exquisite determination of the planet's gravity field and of
  radiometric measurements to probe the deep water abundance.
\item It would be highly desirable to send a probe similar to the
  Galileo probe into Saturn's atmosphere. The comparison of the
  abundance of noble gases would discriminate between different models
  of the enrichment of the giant planets, and the additional
  measurement of key isotopic ratio would provide further tests to
  understand our origins.
\end{itemize}

Clearly, there is a lot of work on the road, but the prospects for a
much improved knowledge of giant planets and their formation are
bright.

\bibliography{geophys_guillot.bib}

\end{document}